\documentclass[
 aps,
 amsmath,amssymb,
reprint,
superscriptaddress,
tightenlines,
floatfix,
nofootinbib,
]{revtex4-1}

\usepackage{graphicx}
\usepackage{dcolumn}
\usepackage{bm}

\usepackage{xcolor}
\usepackage{booktabs}
\usepackage{wasysym}
\usepackage[uncertainty-mode = separate]{siunitx}
\usepackage[utf8]{inputenc}
\usepackage[T1]{fontenc}
\usepackage{mathptmx}
\usepackage{etoolbox}
\usepackage{hyperref}
\usepackage{stmaryrd}

\hypersetup{
    colorlinks = true,
    linkcolor = blue,
    citecolor = blue,
    urlcolor=blue,
}

\newcommand{\blue}[1]{\textcolor{blue}{#1}} 

\newcommand{\black}[1]{\textcolor{black}{#1}} 

\usepackage{tabularx}
\usepackage{appendix}
\usepackage{caption}
\makeatletter
\def\@email#1#2{%
 \endgroup
 \patchcmd{\titleblock@produce}
  {\frontmatter@RRAPformat}
  {\frontmatter@RRAPformat{\produce@RRAP{*#1\href{mailto:#2}{#2}}}\frontmatter@RRAPformat}
  {}{}
}%
\makeatother

\usepackage[font=small,labelfont=bf,format=plain]{caption}

\usepackage{setspace}
  \usepackage{parskip}
\usepackage[compact]{titlesec}         
\titlespacing{\section}{4pt}{4pt}{4pt} 
\AtBeginDocument{
  \setlength\abovedisplayskip{4pt}
  \setlength\belowdisplayskip{4pt}
  }
  
\usepackage[]{caption, subfig}
\usepackage{lineno}
\setcitestyle{square}

\begin{document}
\newcommand{\ra}[1]{\renewcommand{\arraystretch}{#1}}

\title{Resistive diffusion and radiative cooling effects in magnetized oblique shocks}
\author{R. Datta}
\thanks{rdatta@mit.edu}
\affiliation{Plasma Science and Fusion Center, Massachusetts Institute of Technology, MA 02139, Cambridge, USA\looseness=-10000 
}%

\author{E. Neill}%
\affiliation{Plasma Science and Fusion Center, Massachusetts Institute of Technology, MA 02139, Cambridge, USA\looseness=-10000 
}%

\author{E. Freeman}
\affiliation{ 
Laboratory of Plasma Studies, Cornell University, Ithaca, NY 14853, USA 
}%

\author{E. S. Lavine}
\affiliation{ 
Laboratory of Plasma Studies, Cornell University, Ithaca, NY 14853, USA 
}%

\author{S. Chowdhry}%
\affiliation{Plasma Science and Fusion Center, Massachusetts Institute of Technology, MA 02139, Cambridge, USA\looseness=-10000 
}%

\author{L. Horan IV}%
\affiliation{Plasma Science and Fusion Center, Massachusetts Institute of Technology, MA 02139, Cambridge, USA\looseness=-10000 
}%

\author{W. M. Potter}
\affiliation{ 
Laboratory of Plasma Studies, Cornell University, Ithaca, NY 14853, USA 
}%

\author{D. A. Hammer}
\affiliation{ 
Laboratory of Plasma Studies, Cornell University, Ithaca, NY 14853, USA 
}%
\author{B.R. Kusse}
\affiliation{ 
Laboratory of Plasma Studies, Cornell University, Ithaca, NY 14853, USA 
}%

\author{J.D. Hare}%
\thanks{jdhare@cornell.edu}
\affiliation{ 
Laboratory of Plasma Studies, Cornell University, Ithaca, NY 14853, USA 
}%



\begin{abstract}

 Magnetized oblique shocks are of interest in various plasmas, including in astrophysical systems, magneto-inertial confinement fusion experiments, and in aerospace applications. Through experiments on the COBRA pulsed power facility (Cornell University, 1~MA peak current, 100~ns rise time), we investigate oblique shock formation in a system with a magnetic field, and where both radiative cooling and resistive diffusion are important. Compared to previous pulsed power experiments, which have investigated quasi-parallel oblique shocks, here we consider perpendicular-type shocks, which can support magnetic field compression. In our experiments, supersonic, super-Alfvénic, collisional plasma flows, generated using an aluminum exploding wire array, are deflected by angled obstacles to generate oblique shocks. The shocks are imaged using laser shadowgraphy and Mach-Zehnder interferometry, while optical Thomson scattering provides measurements of the flow velocity and temperature. The shocks exhibit shallower shock angles and higher density compression, when compared to canonical Rankine-Hugoniot predictions. These results are best described by a model that includes both resistive diffusion and radiative cooling, consistent with the values of the cooling parameter and the resistive diffusion length in the experiment.

\end{abstract}

\maketitle

\section{Introduction}
\label{sec:intro}


Shocks are an important mechanism of kinetic energy dissipation in many plasmas. In astrophysical systems, shocks facilitate the transfer of energy from galactic interiors to surrounding media. Examples of astrophysical shocks include shocks in core-collapse supernovae and supernova remnants \citep{Chevalier1982,Kifonidis2003}, in protostellar and extragalactic jets \citep{Hartigan1990,Smith2003,Smith2012,Miley1980, Choi2007}, and in weakly-ionized plasmas and molecular clouds \citep{mullan1971structure,draine1980interstellar}. Plasma shocks are also of interest in aerospace applications, for hypersonic travel in plasma environments, or during atmospheric re-entry, where spacecraft interaction with shocked plasma can affect the aerodynamics, thermal transport, and the level of electromagnetic interference \citep{josyula2003governing,semenov2002weakly,keidar2008electromagnetic,shashurin2014laboratory}. In inertial confinement fusion (ICF) \citep{thomas2012drive,atzeni2014shock,craxton2015direct}, shocks drive the compression and heating of the fusion fuel, and in pulsed-power-driven magnetized liner inertial fusion (MagLIF), the driving magnetic pressure shocks and compresses the metal liner imploding under the action of the Lorentz force \citep{gomez2014experimental,ruiz2023exploring,ruiz2023exploring_b}. More recently, there has been a growing interest in applying magnetic fields to ICF implosions \citep{craxton2015direct,walsh2022magnetized,wurden2016magneto,walsh2024resistive}. Magnetic fields can provide several benefits, including reduced thermal transport \citep{walsh2022magnetized} and suppression of deleterious instabilities, which may lead to higher hotspot temperatures and yield amplification \citep{walsh2022magnetized,walsh2025magnetized,chang2011fusion,moody2022increased}. 



Pulsed-power-driven experiments provide an excellent platform for investigating high energy density plasma (HEDP) shocks with embedded magnetic fields \citep{Lebedev2019}. Previous experiments have investigated the structure of magnetized shocks in various geometries, including oblique shocks formed due to the interaction of azimuthally expanding plasma streams \citep{Swadling2013}, normal shocks formed from the interaction of magnetized plasma flows with planar obstacles \citep{Lebedev2014}, and quasi 2-D \citep{Ampleford2010,Bott-Suzuki2015,Burdiak2017,russell2022perpendicular}, and 3-D bow shocks \citep{datta2022structure,datta2022time,datta2024radiatively} generated due to the interaction of plasma flows with blunt obstacles. The inclusion of magnetic fields has been shown to create magnetic precursors upstream of the primary shock generated by planar obstacles \citep{Lebedev2014}, and increase the opening angles of bow shocks by magnetic draping \citep{Burdiak2017}. Ideal magnetohydrodynamic (MHD) shocks that provide simultaneous compression of the density and magnetic field in super-magnetosonic flows have also been predicted by numerical simulations upstream of the current sheet in strongly driven magnetic reconnection \citep{datta2024simulations}, and later visualized experimentally on the Z-machine \citep{datta2024evidence,datta2024thesis}.

 The importance of resistive diffusion has been demonstrated in several pulsed-power-driven shock experiments  \citep{Burdiak2017,russell2022perpendicular,datta2022structure}. Here, the shocked plasma typically has a length scale comparable to the resistive diffusion length, allowing the magnetic field to diffuse significantly, rather than being frozen in the plasma motion \citep{Burdiak2017,russell2022perpendicular,datta2022structure}. Resistive diffusion reduces the importance of magnetic fields, producing plasma shocks closer to the hydrodynamic result \citep{russell2022perpendicular,Burdiak2017,datta2022structure}. Resistive effects can be important in shocks that appear in low temperature, weakly ionized astrophysical and space plasmas. Although resistive effects are considered to be small in ICF implosions, magnetic field diffusion can be important at low temperatures, such as during the early stages of the laser drive, or at the ice-ablator interface \citep{walsh2024resistive}. 

Many of the pulsed-power-driven experiments described above have investigated collisional shocks, given the high density and thus, small collisional mean free path, obtained in these experiments. Some experiments, however, have shown evidence of collisionless physics, such as reflected ions, at early times when the plasma density is low \citep{Lebedev2014}. In astrophysical and space applications, collisionless effects are often important, and the conversion of the pre-shock plasma to the post-shock state is facilitated via two-fluid dispersive effects, as opposed to collisional dissipation. However, collisional shock experiments still provide valuable insight into the fundamental physics of magnetized shocks, since the overall structure of the flow still follows global fluid-like conservation laws \citep{Lebedev2019,borkowski1989two}, and that nonlinear relaxation processes and micro-instabilities can increase the effective collisionality of the medium \citep{Lebedev2019,coburn2022measurement}. 

Another topic of interest in HEDP shock research is plasma radiation. Radiative effects are important in astrophysical, space,  and ICF contexts \citep{Drake2006,Remington1997}. Although the properties of radiative shocks typically depend on the optical thickness of the pre- and post-shock regions \citep{Drake2006,mcclarren2010theory,koenig2006radiative,michaut2007theoretical}, the loss of internal energy via radiative cooling in optically thin systems can result in strong post-shock compression, often termed "catastrophic density collapse" \citep{Drake2006}. Furthermore, radiative cooling can also generate instabilities \citep{field1965thermal,markwick2024cooling}, which have been thought to generate post-shock density perturbations in pulsed-power-driven experiments of planar shocks \citep{merlini2023radiative}. 

The present work expands upon previous pulsed-power-driven magnetized shock research, by investigating magnetized oblique shocks. Oblique shocks form when the upstream velocity vector is not parallel to the shock normal, and provide a partial dissipation of the upstream kinetic energy \citep{goedbloed_keppens_poedts_2010}. They may occur in ICF plasmas due to non-symmetric drive and compression of the fuel \citep{thomas2012drive,walsh2022magnetized}, in aerospace applications from the interaction of hypersonic flow with solid vehicular surfaces, and in astrophysical systems, such as in the "internal working surfaces" of Herbig-Haro jets \citep{Hartigan1990}, and bends of extragalactic jets \citep{Miley1980,Choi2007}. In our experiments, oblique shocks form when supersonic,  super-Alfvénic plasma flows ($M_0 \approx 5.6, \beta \approx 0.5$) generated by an exploding wire array, driven by the COBRA pulsed power machine \citep{greenly20081}, are redirected by angled conducting obstacles. Swadling \textit{et al.} \citep{Swadling2013} have previously investigated the oblique shocks that inherently form in wire arrays due to the azimuthal expansion of the plasma jets streaming from adjacent wire cores. These oblique shocks can be best described as quasi-parallel shocks, where the shock normal is mostly parallel to the magnetic field direction. Parallel shocks are well-known from MHD theory to provide no compression of the magnetic field, and result in purely hydrodynamic solutions \citep{goedbloed_keppens_poedts_2010}. In contrast, the oblique shocks in the present work can be described as "perpendicular" shocks, where the magnetic field is perpendicular to the shock normal, and can support magnetic field compression \citep{goedbloed_keppens_poedts_2010,boyd_sanderson_2003}. Furthermore, previous experiments have also not investigated resistive diffusion and cooling effects in an oblique geometry, and the present work seeks to fill these gaps.

The level of compression in a shock depends on  the polytropic index $\gamma$, which, in high-energy-density (HED) plasmas, may be smaller than the ideal value of $5/3$, because of the contribution of Coulomb interactions, ionization energy, and excitation energy  \citep{Drake2006}. Deviations from the ideal value can cause large changes in compression. Since the shock angle in oblique shocks is sensitive to $\gamma$ \citep{Kundu-2012}, a comparison of shocks for different deflection angles could potentially serve as a diagnostic for the measurement of the polytropic index. Therefore, in addition to investigating the structure of magnetized oblique shocks, we further explore the possibility of inferring $\gamma$ in these oblique shock experiments, which can serve as a proxy for the importance of HED effects, as well as of cooling, as shown later.


The paper is structured as follows. In \blue{Sec.}~\ref{sec:theory}, we present analytical expressions for the shock angle and compression in a simplified representation of the oblique shock in our experiment. In particular, we consider shock formation under the resistive and strongly cooled limits.  In \blue{Sec.}~\ref{sec:methods}, we describe the experimental setup and diagnostics. The experimental results are described in \blue{Sec.}~\ref{sec:results}, while \blue{Sec.}~\ref{sec:discussion} provides a detailed discussion of the shock structure and geometry observed in the experiment, and a comparison with analytical theory, which indicates that the experimentally observed shocks are best described as dominated by radiative cooling and resistive diffusion.

\section{Theoretical Model}
\label{sec:theory}

\begin{figure}[t!]
\includegraphics[page=1,width=0.48\textwidth]{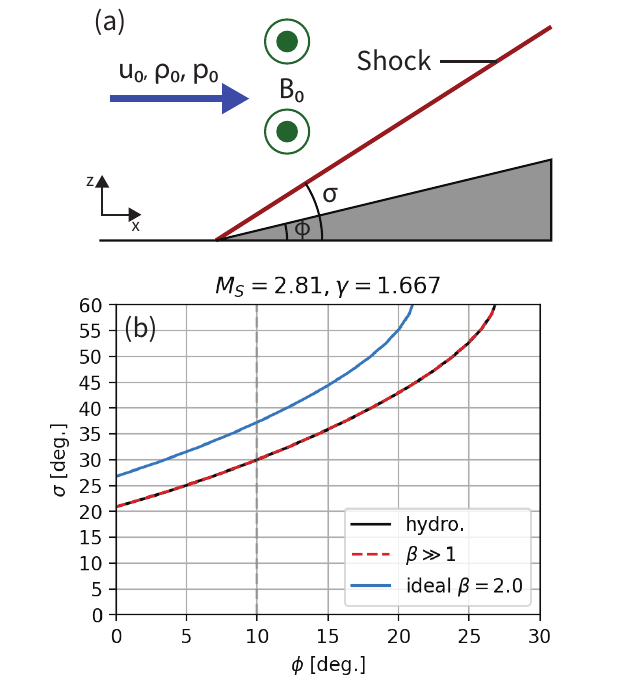}
\centering
\caption{ (a) Oblique shock geometry. A fluid with uniform velocity $u_0 {\bf e_x}$, density $\rho_0$, and pressure $p_0$ is deflected by an angle $\phi$ across an oblique shock that forms at a shock angle $\sigma$. The pre-shock magnetic field ${\bf B_0} = B_0 {\bf e_y}$ points along the out-of-plane direction. (b) The variation of the shock angle $\sigma$ with the deflection angle $\phi$ for an upstream Mach number $M_0 = 2.81$, plasma beta $\beta = 2$, and $\gamma = 5/3$. The large $\beta \gg 1$ solution of \autoref{eq:phi_vs_sigma} is shown using the dashed red line, while the purely hydrodynamic solution (\autoref{eq:phi_vs_sigma_hydro}) is shown using the solid black line.
}
\label{fig:obl_model}
\end{figure}

In this section, we outline a simple model that describes the shock angle $\sigma$ and compression ratios for an oblique shock with an out-of-plane magnetic field. We consider a single-temperature magnetohydrodynamic (MHD) fluid with uniform velocity $u_0 {\bf e_x}$, density $\rho_0$, pressure $p_0$ and constant polytropic index $\gamma$, that is deflected by an angle $\phi$ across an oblique shock that forms at a shock angle $\sigma$, as illustrated in \autoref{fig:obl_model}\blue{a}. This provides a simplified representation of the oblique shock geometry in the experiment (see \blue{Sec.}~\ref{sec:methods}). The normal and tangential vectors to the shock in the $xz$ plane are ${\bf \hat{n}} = \sin \sigma {\bf e_x} - \cos \sigma {\bf e_z}$ and ${\bf \hat{t}} = \cos \sigma {\bf e_x} + \sin \sigma {\bf e_z}$ respectively. The pre-shock magnetic field ${\bf B_0} = B_0 {\bf e_y}$ points along the out-of-plane direction. We denote pre-shock quantities immediately upstream of the shock  with the subscript 0,  while post-shock quantities immediately downstream of the shock are unsubscripted. In addition, the subscripts n and t represent the dot product of a vector quantity with the unit vectors ${\bf n}$ and ${\bf t}$. Here, for simplicity, we have assumed that the upstream velocity is parallel to the $x$-axis; when the pre-shock velocity has a non-zero $z$-component, the shock and deflection angles must be computed relative to the direction of the upstream velocity vector, as described later in \blue{Sec.}~\ref{sec:discussion}. 

When resistive diffusion is important, the Rankine-Hugoniot jump conditions must include the transport of magnetic field and magnetic energy across the shock in the induction and total energy equations respectively \cite{goedbloed_keppens_poedts_2010}, analogous to how the transport of internal energy across the shock front must be accounted for when thermal conduction \citep{lacey1988structure,borkowski1989two} or radiation transport \citep{doss2011oblique} are important.  The resulting Rankine-Hugoniot jump conditions are shown below:

\begin{equation}
\rho u_n = \rho_0 u_{n,0}
\label{eq:continuity}
\end{equation}
\begin{equation}
 B u_n + F = B_0 u_{n,0}
\label{eq:idn}
\end{equation}
\begin{equation}
p+\rho u_{n}^2+\frac{B^2}{2 \mu_0}=p_0+\rho_0 u_{n,0}^2+\frac{B_0^2}{2 \mu_0}
\label{eq:momentum}
\end{equation}
\begin{equation}
\begin{split}
    &\rho_0 u_{n,0}\left(\frac{u_n^2}{2}+\frac{B^2}{\mu_0 \rho}+\frac{\gamma p}{\rho(\gamma-1)}\right) + Q \\
    &=\rho_0 u_{n,0}\left(\frac{u_{n,0}^2}{2}+\frac{B_0^2}{\mu_0 \rho_0}+\frac{\gamma p_0}{\rho_0(\gamma-1)}\right)
\end{split}
\label{eq:energy}
\end{equation}
\begin{equation}
    u_t = u_{t,0}
\end{equation}

 Here, $F = - \bar{\eta}\nabla_n B$ and $Q = -\bar{\eta}\nabla_n(B^2/2\mu_0) = BF/\mu_0$ represent the net transport of magnetic field and magnetic energy from the post-shock to the pre-shock region. Note that for large gradients, diffusive models break down, and the rate of transport saturates at a level set by the thermal velocity of the particles that comprise the fluid \citep{lacey1988structure,borkowski1989two}. These simplified jump conditions describe a class of MHD shocks called perpendicular shocks, where the shock normal is perpendicular to the upstream magnetic field ${\bf B_0 \cdot \hat{n}} = 0$ \citep{goedbloed_keppens_poedts_2010,boyd_sanderson_2003}. 
 
 \textit{Ideal MHD Solution} - In the ideal MHD limit ($F \rightarrow 0$), these shocks are characterized by equal compression ratios for the magnetic field and the mass density $\rho/\rho_0 = u_{n,0}/u_n = B/B_0 \equiv r$. This compression ratio $r$ can be determined from the solution to the quadratic equation \citep{boyd_sanderson_2003}:

\begin{equation}
\begin{split}
2(2-\gamma) r^2+\left[2 \gamma(\beta+1)+\beta \gamma(\gamma-1) M_{n,0}^2\right] r \\
- \beta \gamma(\gamma+1) M_{n,0}^2=0
\end{split}
\label{eq:perp_eqn}
\end{equation}

Where $M_{n,0} = \sqrt{ \rho_0 u_{n,0}^2/\gamma p_0}$ is the pre-shock sonic Mach number, calculated using the upstream velocity $u_{n,0}$ normal to the shock, and $\beta = 2 \mu_0 p_0/ B_0^2$ is the upstream plasma beta.

To relate the shock geometry to the compression ratio $r$, we decompose the upstream velocity ${\bf u_0}$ into its normal and tangential components as $u_{0,n} = u_0 \sin \sigma$ and $u_{0,t} = u_0  \cos \sigma$. Similarly, the post-shock velocity components are $u_{n} = u \sin (\sigma -\phi)$ and $u_{t} = u \cos (\sigma -\phi)$. Then, the compression ratio $r \equiv u_{0,n} / u_{n}$ becomes:

\begin{equation}
r = \frac{u_{0} \sin \sigma}{u \sin (\sigma-\phi)} = \frac{u_{0,t} \tan \sigma}{u_{t} \tan (\sigma - \phi)} = \frac{\tan \sigma  (1+\tan \sigma \tan \phi)}{\tan \sigma-\tan \phi}
\end{equation}

Here, we have used the jump condition $u_{t} = u_{0,t}$ and the trigonometric identity $\tan (a+b)=(\tan a+\tan b)/(1-\tan a \tan b)$ to simplify the expression. Rewriting the above equation, we get:

\begin{equation}
\tan \phi=\left[\frac{\tan \sigma(r-1)}{r+\tan ^2 \sigma}\right]
\label{eq:phi_vs_sigma}
\end{equation}

Here, the compression ratio $r$ is a function of the upstream Mach number $M_{0,n} = M_0 \sin \sigma$ (normal to the shock), the plasma beta $\beta$, and the polytropic index $\gamma$, obtained from the solution of \autoref{eq:perp_eqn}.

As an example, the blue curve in \autoref{fig:obl_model}\blue{b} shows the (weak branch \citep{Kundu-2012}) solution of the shock angle $\sigma$ as a function of the deflection angle $\phi$ for an upstream Mach number $M_0 = 2.81$, plasma beta $\beta = 2$, and $\gamma = 5/3$, calculated from the solutions of \autoref{eq:perp_eqn} \blue{and} \ref{eq:phi_vs_sigma}, while the black curve represents the well-known purely hydrodynamic solution \citep{Kundu-2012}:

\begin{equation}
    \tan \phi=2 \cot \sigma \frac{M_0^2 \sin ^2 \sigma-1}{M_0^2(\gamma+\cos 2 \sigma)+2}
    \label{eq:phi_vs_sigma_hydro}
\end{equation}

For a given deflection angle $\phi$, the ideal MHD shock is expected to form at a larger angle than in the hydrodynamic case. In the large $\beta \gg 1$ case, the result of \autoref{eq:phi_vs_sigma} approaches the hydrodynamic solution (\autoref{eq:phi_vs_sigma_hydro}) as expected. 

To demonstrate the analytical result shown in \autoref{fig:obl_model}\blue{b}, we perform a simple two-dimensional ($xz$-plane) nonlinear MHD simulation, where a flow ($M_0 = 2.81, \beta = 2$) with an out-of-plane magnetic field is deflected by $\phi = \SI{10}{\degree}$ by a conducting wedge-shaped obstacle. Full details on the simulation setup and results are provided in \autoref{appendixA}. We briefly summarize the numerical results here. The simulated shock forms at an angle of $\sigma =\SI{37}{\degree}$, which agrees with the theoretical prediction shown in \autoref{fig:obl_model}\blue{b}. The density and magnetic field exhibit equal compression ($r = 1.47$) across the shock front, consistent with the theoretical prediction calculated using \autoref{eq:perp_eqn}. The pressure compression is $R = 1.96$, again consistent with the value expected from analytical theory.  In the hydrodynamic ($\beta \rightarrow \infty$) case, the simulated shock forms at $\sigma =\SI{30}{\degree}$, consistent with the theoretical prediction shown in \autoref{fig:obl_model}\blue{b}. The density and pressure compression ratios are $r = 1.59$ and $R = 2.22$, which agree with the values expected from analytical hydrodynamic theory.

\textit{Effect of Resistive Diffusion –} When resistive effects are important, we must retain the transport terms $F$ and $Q$ in the jump conditions (\autoref{eq:continuity}-\ref{eq:energy}). While jump conditions have been derived under conditions of radiative or conductive transport \citep{doss2011oblique,lacey1988structure,borkowski1989two}, we extend the derivation to include magnetic field transport. We express the jump conditions in dimensionless form by normalizing the relevant quantities as $\mathcal{N} \equiv \rho_0/\rho = u_n/u_{0,n}$, $b \equiv B/(u_{0,n}\sqrt{\mu_0 \rho_0})$, $\tau \equiv (p/\rho)/u_{0,n}^2$, and $f = F/(u_{0,n}^2\sqrt{\mu_0 \rho_0})$. Furthermore, assuming $\gamma = 5/3$, we get:

\begin{equation}
\begin{aligned}
    &f+b \mathcal{N}=b_0 \\
    &\tau =  \mathcal{N} \left(\frac{b_0^2-b^2}{2}-\mathcal{N}+\tau_0+1\right)\\
    & 2 b^2 \mathcal{N}+2 f b+\mathcal{N}^2+5 \tau=2 b_0^2+5 \tau_0+1
\end{aligned} 
\label{eq:normalized_eqns}
\end{equation}

Lastly, eliminating $\tau$ and $b$  from these equations gives us a single equation for the inverse density jump $\mathcal{N} = r^{-1}$ in terms of $\tau_0, b_0$ and $f$:  \footnote{For arbitrary $\gamma$, the jump condition \autoref{eq:final_eqn} can be written as  \\ $(\gamma+1) \mathcal{N}^3-2 \gamma \left(\frac{b_0^2}{2}+ \tau_0+1\right) \mathcal{N}^2-\left(2 b_0^2 (1-\gamma)-2\gamma \tau_0+1 - \gamma\right) \mathcal{N}+b_0^2(2-\gamma)-2 b_0 f+\gamma f^2=0$}

\begin{equation}
\begin{split}
  &4 \mathcal{N}^3-\left(\frac{5 b_0^2}{2}+5 \tau_0+5\right) \mathcal{N}^2\\&+\left(2 b_0^2+5 \tau_0+1\right) \mathcal{N}+\frac{b_0^2}{2}-3 b_0 f+\frac{5 f^2}{2}=0
    \end{split}
    \label{eq:final_eqn}
\end{equation}

Note that $\tau_0$ is related to the upstream sonic Mach number as $\tau_0 = (\gamma M_{0,n}^2)^{-1}$, while $b_0$ is related to the upstream Alfvénic Mach number $b_0 = M_{A,n}^{-1}$. For $b > b_0$, the induction equation shown in \autoref{eq:normalized_eqns} provides the condition $f < b_0(1-\mathcal{N})$. Moreover, since $0 < \mathcal{N} \leq 1$, $f$ cannot exceed $b_0 \equiv 1/M_{A,n}$, and the transport must thus be sub-Alfvénic. 

For $f = 0$, we recover the ideal MHD solution $b = b_0/\mathcal{N}$, where $\mathcal{N}$ is obtained from the solution of the following equation in the range $0 < \mathcal{N} < 1$:
\begin{equation}
    \mathcal{N}^2\left(\frac{5 b_0^2}{2}+5 \tau_0+5\right)-\mathcal{N}\left(2 b_0^2+5 \tau_0+1\right)-\frac{b_0^2}{2}-4 \mathcal{N}^3=0
\end{equation}
This equation, after eliminating the $\mathcal{N} = 1$ solution, can be shown to be equivalent to the quadratic density jump equation shown earlier (\autoref{eq:perp_eqn}). Similarly, setting $b_0 = 0$ enables us to recover the hydrodynamic solution $\mathcal{N} = (1+5\tau_0)/4$. 

We consider the limiting case where magnetic transport is large enough to cause the magnetic field immediately upstream and downstream of the shock to be roughly equal $b_0 \approx b$. Formally, this will happen when the $f \approx (1-\mathcal{N})b \equiv f_{max}$, as seen from the induction equation in \autoref{eq:normalized_eqns}. Substituting this result into \autoref{eq:final_eqn} removes any dependence of the solution on the magnetic field $b$ as shown below, and the solution to which is identical to the hydrodynamic result $\mathcal{N}_{hydro} = (1+5\tau_0)/4$. 
\begin{equation}
    4\mathcal{N}^2+\left(-5 \tau_0-5\right) \mathcal{N}+5 \tau_0+1 = 0
\end{equation}
To demonstrate this result, we repeat the 2-D simulation ($M_0 = 2.81, \beta = 2, \phi =\SI{10}{\degree}$) reported previously, but with a large value of magnetic diffusivity, providing a magnetic Reynolds number $R_M = u_0L/\bar{\eta} \approx 1$. Here, resistive diffusion results in a magnetic precursor ahead of the density jump. The change in the magnetic field across the shock is small, and as expected, the shock angle ($\sigma = \SI{30}{\degree}$) and density and pressure jump ($r = 1.59, \, R = 2.22$) immediately across the shock are consistent with the hydrodynamic solution. Further details of this simulation are again provided in \autoref{appendixA}. Therefore, when resistive effects are important, we expect the shock angle to decrease and move closer to the hydrodynamic solution, consistent with experimental observations of shock formation in resistive plasmas \citep{Burdiak2017,datta2022structure}. For intermediate values of $0 < f < (1-\mathcal{N})b_0$, solutions that lie between the hydro- and ideal MHD results are expected.

\textit{Effects of Radiative Cooling –}  Several authors have explored how cooling affects shock formation \citep{lacey1988structure,borkowski1989two,creasey2011numerical,draine1980interstellar,Drake2006}. The loss of thermal energy from optically thin systems is known to provide unbounded density compression, only limited by some mechanism that sets the minimum temperature or compression achievable within the system \citep{Drake2006}. In radiative shocks, radiative precursors that heat the pre-shock plasma can propagate upstream due to re-absorption of emission from the hot post-shock region \citep{doss2011oblique,Drake2006,mcclarren2010theory,koenig2006radiative,michaut2007theoretical}. As described later in \blue{Sec.}~\ref{sec:discussion}, we expect the majority of the radiation to leave the system in our experiment, and therefore, neglect the impact of radiation transport in this analysis for simplicity.

For a hydrodynamic fluid (with $\gamma = 5/3$), assuming an idealized system where the radiative cooling rate is non-zero only for temperatures greater than some critical value $T_c$, the final post-shock state will exhibit an inverse density jump of $\mathcal{N}_c = (1+\tau_0)/2-[1/4(1+\tau_0)^2-\tau_c]^{1/2}$ \citep{lacey1988structure,borkowski1989two}. For the special case where the critical temperature is equal to the pre-shock temperature $T_c = T_0$, the system is driven to thermal equilibrium by radiative cooling. In this idealized isothermal limit, where the density and pressure compression ratios are now equal, solving the Rankine-Hugoniot jump conditions shows that the density jump in the final state is $r = R = \tau_0^{-1}$ in the hydrodynamic case, and $r = R = 1/2\left[-(\beta+1)+\sqrt{(\beta+1)^2+4\tau_0^{-1}\beta}\right]$ in the ideal MHD case, where $\tau_0^{-1} \equiv \gamma M_{n,0}^2 = \rho_0u_{0,n}^2/p_0$. In both cases, this describes an equivalent system with $\gamma_\text{eff} \rightarrow 1$. Indeed, when we repeat our 2-D simulations with radiative cooling ($t_\text{cool}^{-1}/t_{\text{hydro}}^{-1} \approx 2.5, T_c = T_0$), the shock angle decreases upon onset of cooling until a steady solution is reached (see \autoref{appendixA}). The temperature rises at the shock front and then falls rapidly behind it due to radiative cooling. In both the hydro- and ideal MHD cases, the simulated shock angle and compression ratios match the analytical prediction obtained for $\gamma_\text{eff} \rightarrow 1$ and $M_\text{eff} = u_0 / \sqrt{p_0/\rho_0} \approx 3.62$. 

Therefore, while an out-of-plane magnetic field is expected to increase the oblique shock angle when compared to the hydrodynamic solution, resistive diffusion and radiative cooling can cause the shock to form at a shallower angle. The models presented in this section will inform the analysis of our experiment, which is described next.

\begin{figure}[t!]
\includegraphics[page=1,width=0.48\textwidth]{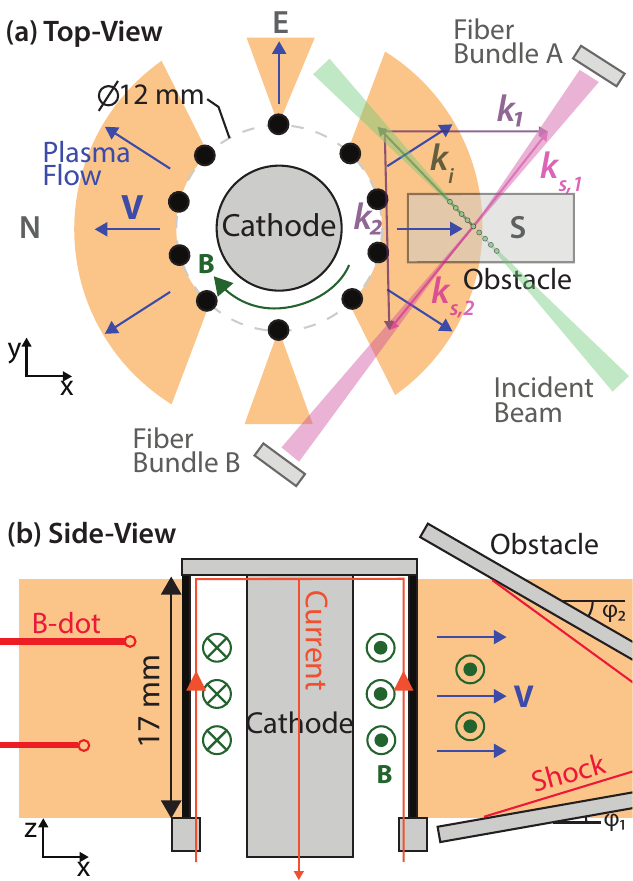}
\centering
\caption{ (a) Top view of the load hardware, showing the exploding wire array and obstacle geometry. The wire array consists of 10, \SI{30}{\micro\meter} diameter aluminum wires, and
has a wire-cathode separation of roughly \SI{3}{\milli\meter}. Geometry for optical Thomson scattering is also shown. The incident beam (10~J, Nd-YLF laser, 526.5~nm) with wavevector ${\bf k_i}$ is focused into a  diameter of about $\SI{250}{\micro\meter}$. The incident wavevector lies entirely in the $xy-$plane, and the focal point is at about $x = \SI{8}{\milli\meter}$ from the wires, and roughly $\SI{8.5}{\milli\meter}$ above the base of the wires. Collection angles are $\pm \SI{90}{\degree}$ to the probe beam. (b) Side view of the load hardware, showing obstacles that redirect the flow generating oblique shocks.
}
\label{fig:setup}
\end{figure}

\section{Experimental Setup}
\label{sec:methods}
\subsection{Load Hardware}

The load hardware is an exploding wire array (height = $\SI{17}{\milli\meter}$, diameter = $\SI{12}{\milli\meter}$), comprising a cylindrical cage of 10, \SI{30}{\micro\meter} diameter aluminum wires around a central cathode, as shown in \autoref{fig:setup} \citep{Lebedev2019}. The wire-cathode separation is roughly \SI{3}{\milli\meter}. The wires are positioned such that four wires,  with a $\SI{30}{\degree}$ angular separation, are on the north and south sides, and one on the east and west sides each, as shown in \autoref{fig:setup}\blue{a}. The inter-wire gap is aligned with the obstacle, as illustrated in \autoref{fig:setup}\blue{a}. The number of wires is chosen to prevent mass depletion of the wire array over the duration the experiment, as calculated via the rocket model \citep{Lebedev2001}. The wire array is driven by a $\sim$1~MA peak, $\sim$100~ns rise time current pulse generated by the COBRA generator \citep{greenly20081}. When current flows through the wires, coronal plasma, which forms around the wire cores, is accelerated radially outwards by the global ${\bf j \times B}$ force in the wire-cathode gap. This results in radially diverging supersonic plasma flows, which advect magnetic field from the cathode-wire gap into the flow region \citep{Lebedev2014,Lebedev2019}. At the $y=\SI{0}{\milli\meter}$ mid-plane, the magnetic field vector points in the out-of-plane $y$-direction, as shown in \autoref{fig:setup}\blue{b}. The overmassed wire array provides continuous sustained plasma flows from the ablation of the wire material during the experiment \citep{datta2023plasma}. Optical imaging of the wire array during the shot (not shown here) indicates that the array begins to explode due to mass depletion around $200\text{-}\SI{210}{\nano\second}$, which is after the time of interest in these experiments. 

The flows are deflected by planar aluminum obstacles placed in the flow, as shown in \autoref{fig:setup}\blue{b}. The obstacles are about $\SI{0.5}{\milli\meter}$ thick, and $\SI{3}{\milli\meter}$ wide in the $y$-direction. The obstacles form angles of roughly $\phi_1 = \SI{13}{\degree}$ and $\phi_2 = \SI{31}{\degree}$ with respect to the horizontal, and the angular deflection of the flow generates oblique plasma shocks. The width and position of the obstacles are chosen so that they interact only with the plasma from the inter-wire gap, making it easier to interpret our results. 

Exploding wire arrays have been previously used in various shock experiments \citep{Burdiak2017,datta2022structure,russell2022perpendicular}, and offer advantages such as good diagnostic access and unimpeded flow into a large volume. These wire arrays have been well characterized on 1~MA university-scale machines, such as MAGPIE \citep{Lebedev2019}. Exploding wire arrays with aluminum wires typically generate collisional plasma flows $\lambda_{ii} \approx \SI{1}{\nano \meter}$ with characteristic density $n_e \approx 1-\SI{10e17}{\per \centi \meter \cubed}$, temperature $T_e \approx 5-\SI{10}{\electronvolt}$, velocity $V \approx \SI{100}{\kilo \meter \per \second}$, and sonic Mach number $M_0 > 5$ \citep{Suttle_2019,datta2022structure,datta2023plasma}. We also note that due to the high density of the plasma, the plasma sheath width ($\lambda_D\approx \SI{50}{\nano\meter}$) at the obstacle-plasma interface is expected to be small. Because of the large driving magnetic pressure in the wire-cathode gap, the flow velocity in the pre-shock region is mostly directed radially outwards from the wires. However, above and below the wire array, the plasma is free to expand due to its pressure into the vacuum region. This results in a small pressure-gradient-driven $z$-component of the velocity. As we note later in \blue{Sec.}~\ref{sec:discussion}, this may affect the upstream velocity direction, and thus, the angle of shock formation in the experiment.



\begin{figure*}[t!]
\includegraphics[page=3,width=1.0\textwidth]{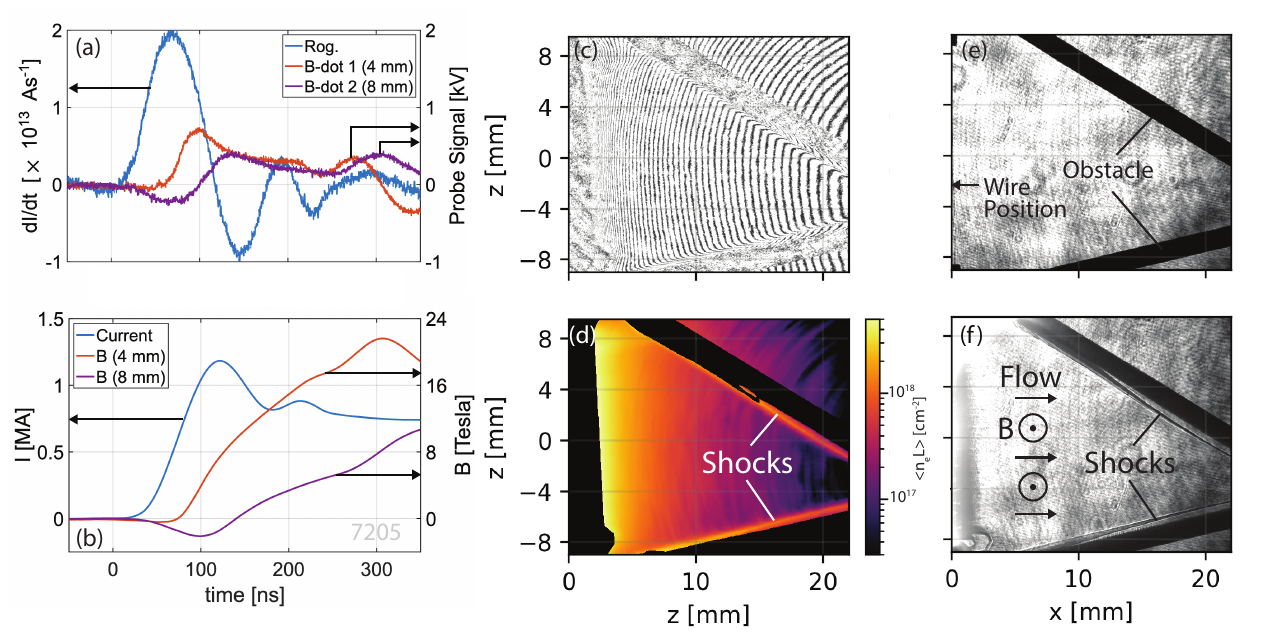}
\centering
\caption{ (a) The Rogowski and B-dot probe voltage signals. (b) Current and magnetic field measurements obtained from numerical integration of the signals shown in (a). (c) Shot interferogram recorded at \SI{170}{\nano\second} after current start. (d) Line-integrated electron density calculated from (c) using MAGIC2 \citep{Swadling2013}. Shocks appear as regions of enhanced $\langle n_eL\rangle$ near the obstacles. Regions near the obstacles and wires, where the fringes are difficult to trace, have been masked out.(d) Pre-shot shadowgraph showing the obstacle position. (e) Shot shadowgraph recorded at \SI{170}{\nano\second} after current start. 
}
\label{fig:results}
\end{figure*}

\subsection{Diagnostics}

A Rogowski coil in the magnetically insulated transmission line of the COBRA machine monitors the current delivered to the load \citep{greenly20081}. The Rogowski coil provides a voltage signal, which is then integrated numerically and combined with a known calibration factor to determine the delivered current.

B-dot probes are used to measure the time-resolved magnetic field advected by the plasma. Two probes are positioned at $\SI{4}{\milli\meter}$ and $\SI{8}{\milli\meter}$ from the wires on the side of the array opposite that facing the obstacle (see \autoref{fig:setup}\blue{b}). Each probe consists of a single-turn \SI{1}{\milli\meter} diameter loop of wire made from connecting the inner and outer conductors of a coaxial cable \citep{byvank2017applied,datta2022time}. We insulate the probe tips with polyamide film. The probes are pre-calibrated, and their voltage signals are integrated numerically to determine the advected magnetic field. The effective loop areas of the probes are $A_{1} \approx \SI{3.9}{\milli\meter\squared}$ and $A_{2} \approx \SI{5.2}{\milli\meter\squared}$ respectively. Uncertainties in the measured probe area determined during pre-calibration are $<1\%$.

We visualize the shocks using laser shadowgraphy and Mach-Zehnder interferometry. The shadowgraphy and interferometry diagnostics share a probe beam ($\SI{100}{\milli \joule}$, $\SI{532}{\nano \meter}$ Nd:YAG laser with $\SI{150}{\pico \second}$ pulse width) which provides a side-on view ($xz$ plane) of the experimental setup with a roughly \SI{25}{\milli\meter} diameter field of view. A pre-shot shadowgraphy image showing the obstacle and wire positions is shown in \autoref{fig:results}\blue{e}.  As the probe beam propagates through the plasma, electron density gradients introduce deflections in the beam, which result in intensity variations in the shadowgraph. We split off and combine the probe beam with the reference beam to record the interferometry image. The plasma also introduces a phase in the probe beam, proportional to the line-integrated electron density. The interferometry field of view includes a region with no plasma, providing an undistorted zero phase shift reference region where we set the electron density to 0. We trace the fringes by hand and post-process the traced interferogram in MAGIC2 to calculate the line-integrated electron density \citep{hare2019two}. The shadowgraphs and the interferograms are recorded simultaneously using Canon EOS DIGITAL REBEL XS cameras. The focal plane of the imaging system is located at the $y=\SI{0}{\milli\meter}$ mid-plane for interferometry, while it is at $y\approx\SI{-5}{\milli\meter}$ for the shadowgraphy, in order to boost the shadowgraphy signal. Both diagnostics record 1 frame per experimental shot.


We make spatially-resolved measurements of the flow velocity $V$ and ion $T_i$ and electron temperatures ${\bar Z}T_e$ using the collective ion feature of optical Thompson scattering (OTS) \citep{rocco2018time,lavine2022measurements}. The geometry for OTS is shown in \autoref{fig:setup}\blue{a}. The incident beam (10~J, Nd-YLF laser, 526.5~nm) with wavevector ${\bf k_i}$ is focused into a  diameter of about $\SI{250}{\micro\meter}$. The incident wavevector lies entirely in the $xy-$plane, and the focal point is at about $x = \SI{8}{\milli\meter}$ from the wires, and roughly $\SI{8.5}{\milli\meter}$ above the base of the wires. We collect the scattered light simultaneously along two directions $\mathbf{k_{s,1}}$ and $\mathbf{k_{s,2}}$, at collection angles of $\pm\SI{90}{\degree}$ with respect to the probe beam. The scattering vectors ${\bf k_{1,2}}$ are shown in \autoref{fig:setup}\blue{a}. The scattered light is collected by multimode fiber bundles, after focusing by a lens that provides a $M = 2$  magnification of the collection volumes. The fiber bundles contain 10 optical fibers, each with a \SI{100}{\micro\meter} diameter core and \SI{20}{\micro\meter} cladding. The fiber-to-fiber separation is \SI{125}{\micro\meter}. The diameter of the collection volume $\approx \SI{200}{\micro\meter}$ is set by the sizes of the focal spot and the magnified image of the optical fiber. The fiber output is coupled to a 750~mm Czerny–Turner spectrometer with a 2400~l/mm grating, providing a spectral resolution of $\SI{0.2}{\angstrom}$. The OTS diagnostic records scattered light once per shot; the gate time is set by the incident beam's temporal profile, which is Gaussian with a full width at half maximum of 2.3~ns.


\begin{figure}[t!]
\includegraphics[page=4,width=0.48\textwidth]{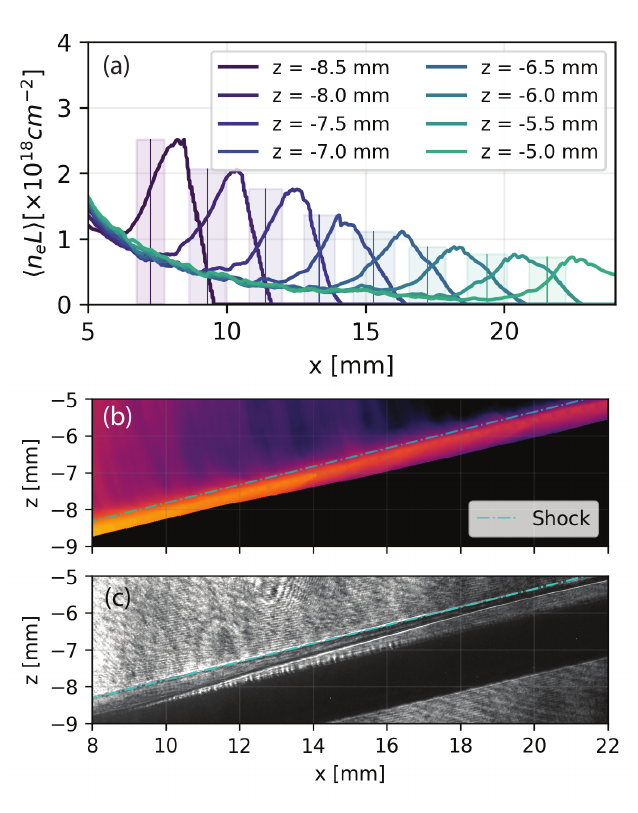}
\centering
\caption{ (a) Variation of $\langle n_eL \rangle$ along the $x$-direction at the bottom shock for different values of the axial position $z$. The vertical lines show the detected shock front and the shaded regions show the associated uncertainty (here, equal to $0.5\times$FWHM of the transition).  (b) A magnified image of the $\langle n_e L \rangle$ showing the density enhancement at the bottom shock. (c) A magnified image of the shadowgraph. The cyan dashed line shows the detected shock front.
}
\label{fig:neL_lineouts}
\end{figure}

\section{Experimental Results}
\label{sec:results}

Multiple experimental shots were performed using the platform described above. Each shot provided repeatable results; here, we describe one such representative shot in detail.

\autoref{fig:results}\blue{(a-b)} show the Rogowski signal and the integrated current delivered during the experiment, respectively. In this shot, COBRA delivered a peak current of $\SI{1.2}{\mega\ampere}$, with a rise time of about \SI{120}{\nano\second}.  Magnetic field measurements from the B-dot probes are shown in \autoref{fig:results}\blue{(a-b)}. The voltage signals from the probes are displaced in time, consistent with the advection of the magnetic field by the plasma flow. From the delay in the probe signals, we estimate an average flow velocity of about $\SI{100}{\kilo\meter\per\second}$. At \SI{170}{\nano\second}, the magnitude of the field is about $\SI{12.5}{\tesla}$ at \SI{4}{\milli\meter}, and $\SI{2.2}{\tesla}$ at \SI{8}{\milli\meter}. We note the probe at 8~mm demonstrates an initial negative voltage before the signal begins to rise. The reason for this is unclear, but may indicate the coupling of the probe with electrostatic voltages at this time.


\autoref{fig:results}\blue{(c-f)} show the interferometry and shadowgraphy images recorded at \SI{170}{\nano\second} after current start. Here, we use a coordinate system such that $x = \SI{0}{\milli\meter}$ represents the wire position, and $z = \SI{0}{\milli\meter}$ is halfway between the anode base and top cathode of the wire array. The interferogram and the shadowgraph show well-defined shocks near both obstacles. In the interferogam (\autoref{fig:results}\blue{c}), the phase shift introduced by the shocks distorts the fringe pattern; this corresponds to enhanced line-integrated electron density $\langle n_eL \rangle$ at the shock front, as observed in the processed $\langle n_eL \rangle$ map shown in \autoref{fig:results}\blue{d}. 

\begin{figure}[b!]
\includegraphics[page=8,width=0.48\textwidth]{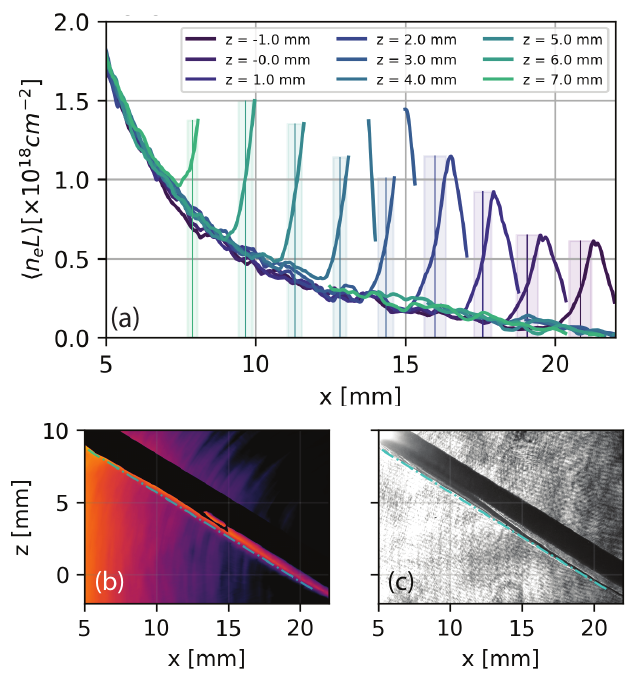}
\centering
\caption{ (a) Variation of $\langle n_eL \rangle$ along the $x$-direction at the top shock for different values of the axial position $z$. The vertical lines show the detected shock front and the shaded regions show the associated uncertainty (here, equal to $0.5\times$FWHM of the transition). (b) A magnified image of the $\langle n_e L \rangle$ showing the density enhancement at the top shock. (c) A magnified image of the shadowgraph. The cyan dashed line shows the detected shock front.
}
\label{fig:neL_lineouts_top}
\end{figure}

\autoref{fig:neL_lineouts}\blue{a} shows the variation of $\langle n_eL \rangle$ along the $x$-direction at the bottom shock for different values of the axial position $z$. As expected, in the pre-shock region, the line-integrated density is roughly uniform in $z$, and falls with distance from the wires, consistent with time-of-flight effects \citep{Lebedev2001}. At the shock front, $\langle n_eL \rangle$ rises abruptly, before falling again in the post-shock region. The peak value of $\langle n_e L\rangle$, as well as the post-shock gradient in $\langle n_e L\rangle$, decrease with distance from the wires. Similar trends in the line-integrated electron density are also observed for the shock at the top obstacle, as shown in \autoref{fig:neL_lineouts_top}\blue{a}. Here, we observe very steep $\langle n_e L\rangle$ gradients close to the wires, and the gradients become smaller further away. We are unable to fully resolve the change in $\langle n_eL \rangle$ across the top shock close to the wires, because the fringes become too close to trace properly (see \autoref{fig:neL_lineouts_top}).

In the shadowgraph (\autoref{fig:results}\blue{f}), the shocks appear as dark regions of low intensity, consistent with the deflection of the probing beam away from these regions. The dark band is symmetrically flanked by thin bright regions, consistent with the focusing of light into these regions. This is seen more clearly in \autoref{fig:neL_lineouts}\blue{c} and \autoref{fig:neL_lineouts_top}\blue{c}, which show magnified images of the shocks. The dark bands also become narrower with distance from the wires at both the top and bottom shocks. The intensity distribution observed in the shadowgraph is largely consistent with the line-integrated electron density gradients measured by interferometry – that is, the peaked $\langle n_e L\rangle$ at the shock front acts as a diverging lens deflecting light away from this region. The narrowing of the dark band with $x$ is also consistent with the decreasing post-shock $\langle n_e L\rangle$ gradient further away from the wires, as observed in \autoref{fig:neL_lineouts}\blue{a} and \autoref{fig:neL_lineouts_top}\blue{a}. Simple ray tracing calculations performed using the experimentally observed  $\langle n_e L\rangle$ distribution confirm this result.

The shadowgraph also shows modulations on the surfaces of the obstacles. These modulations are clearly visible between $\SI{10}{\milli\meter}< x < \SI{17}{\milli\meter}$ in \autoref{fig:neL_lineouts}\blue{c}, exhibiting a wavelength of about $0.1\text{-}\SI{0.15}{\milli\meter}$ and amplitude $<\SI{0.25}{\milli\meter}$. The modulations could result from electro-thermal instability (ETI) growth due to strong surface currents, as discussed later in \blue{Sec.}\ref{sec:discussion}. Notably, these modulations appear in the shadowgraph, but remain unresolved in the interferogram. 



\begin{figure}[t!]
\includegraphics[page=5,width=0.48\textwidth]{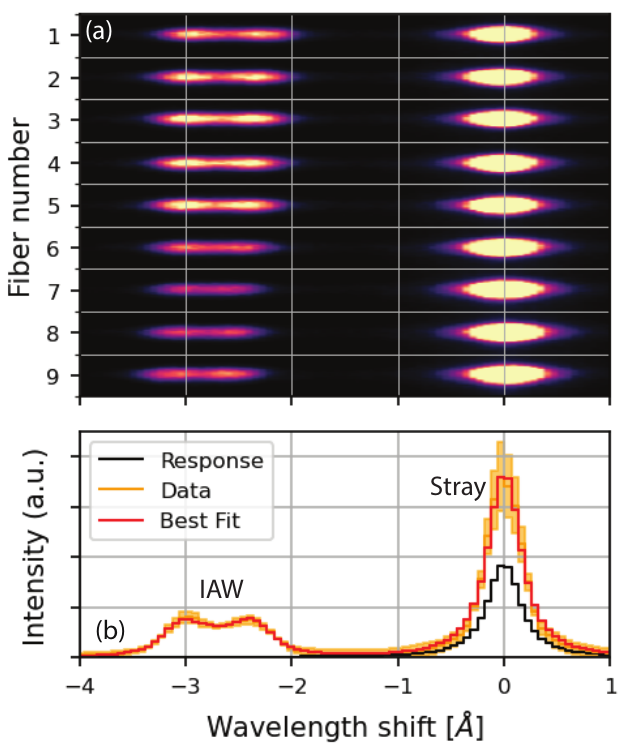}
\centering
\caption{ (a) The raw OTS spectrogram from fiber bundle A recorded at 170~ns. (b) Measured OTS spectrum and least-squares fit for the central fiber (fiber number = 5).
}
\label{fig:OTS_result}
\end{figure}

\autoref{fig:OTS_result}\blue{a} shows the raw OTS spectrogram from fiber bundle A recorded at 170~ns. Reflections in the vacuum chamber result in stray light centered at $\Delta \lambda \equiv \lambda_s - \lambda_i = \SI{0}{\angstrom}$. The Doppler-shifted scattered light appears in the $\Delta \lambda < \SI{0}{\angstrom}$ region for fiber bundle A. The shift increases with distance from the wires, indicating an increase in the flow velocity. For fiber bundle B, the near perpendicular scattering vector ${\bf k_2}$ relative to the expected velocity ${\bf V} \approx V {\bf e_x}$ (see \autoref{fig:setup}\blue{a}) makes the Doppler shift small ${\bf V \cdot k_2} \approx 0$. Thus, the scattered light is not resolved due to the strong stray light signal at $\Delta \lambda = \SI{0}{\angstrom}$, and we therefore focus on results from bundle A.

The scattered spectra exhibit two well-separated peaks, consistent with ion-acoustic peaks in the collective regime \citep{Froula2006,Suttle2021}. The peak separation remains roughly constant, indicating that $\bar{Z}T_e$ does not change significantly over the spatial extent (roughly 1~mm) of the fibers. \autoref{fig:OTS_result}\blue{b} shows the OTS spectra recorded by the central fiber ($N = 5$), which is at roughly $\SI{8}{\milli\meter}$ from the wires. We make quantitative estimates of the flow velocity $V$, electron temperature $\bar{Z}T_e$ and ion temperature $T_i$ through least-squares fitting of synthetic OTS spectra to the observed data \citep{Suttle2021,Hare2018}, which is shown by the red curve in \autoref{fig:OTS_result}\blue{b} for the central fiber. We use measurements of the electron density $n_e$ from the interferometry diagnostic to constrain the fit, and convolve the theoretical spectrum with the spectrometer instrument response function (Gaussian, s.d. = \SI{0.2}{\angstrom}) to generate the synthetic spectra. Here, the value of $n_e$ is estimated from an Abel inversion of the line-integrated electron density at $z = \SI{-2}{\milli\meter}$, where the interferometry measurements are unaffected by the shocks (see \autoref{fig:results}\blue{d
}). 3-D MHD simulations of the wire array geometry in the experiment with the measured COBRA current pulse, performed using the code GORGON \citep{Chittenden2004,Ciardi2007}, indicate that the Abel inverted density is similar to the actual electron density within about 40$\%$, despite the azimuthal asymmetry of the electron density distribution. Furthermore, we note that the values of $ZT_e$, $T_i$ and $V$ determined from the OTS ion feature are relatively insensitive to changes in $n_e$ of this magnitude.

For the central fiber, the ion and electron temperatures are roughly $T_i \approx 8\pm\SI{2}{\electronvolt}$ and $T_e \approx 14\pm\SI{1}{\electronvolt}$, the average ionization is $\bar{Z} \approx 3.9$, and the flow velocity is about $V \approx 110\pm\SI{1}{\kilo\meter\per\second}$ (see \autoref{fig:OTS_result}\blue{b}). Here, we have used the nLTE (non-local thermodynamic equilibrium) model SpK \citep{crilly2023spk} to separate the electron temperature and ionization of the plasma.  The sonic Mach number is $M_0  = V / C_S \approx 5.6$, where $C_S \equiv \sqrt{\gamma p/\rho}$ is the sound speed, $p = n_e T_e + n_i T_i$ is the thermal pressure, and $\rho$ is the mass density. Here, we have used the ideal gas value of $\gamma = 5/3$ in calculating $M_0$. The values of the plasma quantities obtained from OTS exhibit some variation across the spatial extent (roughly \SI{1}{\milli\meter}) covered by the other fibers. The ion and electron temperatures vary between $T_e \approx 11\text{-}\SI{14.5}{\electronvolt}$ and $T_i \approx 6\text{-}\SI{9}{\electronvolt}$ respectively, while the flow velocity increases from $V \approx \SI{105}{\kilo\meter\per\second}$ to roughly $V \approx \SI{117}{\kilo\meter\per\second}$ with distance from the wires. The resulting range in the sonic Mach number is $M_0 \approx 5.3-7.7$. Finally, combining the magnetic field measured at $\SI{8}{\milli\meter}$ with the OTS measurements, we estimate a plasma beta of about $\beta \equiv p / [B^2/(2\mu_0)] \approx 0.5$ in the pre-shock region, while the Alfvén Mach number is about $M_A \approx 3.6$.

The relevant plasma quantities in the pre-shock region determined from the experimental measurements and their associated ranges are summarized in \autoref{tab:table}. We report these parameters for $x=\SI{8}{\milli\meter}$ and $t =\SI{170}{\nano\second}$, for which we have simultaneous measurements from all diagnostics. We reiterate that we do not expect significant variation in these quantities along the $z$-direction \citep{Lebedev2019}. Moreover, although quantities such as density and velocity change with distance from the wires in an exploding wire array, previous work has shown that the sonic and Alfvénic Mach numbers remain roughly constant  \citep{Burdiak2017,Suttle_2019}.


\begin{table}\centering
\ra{1.3}
\caption{
A summary of measured quantities in the unshocked plasma region at about $\SI{8}{\milli\meter}$ from the wires, for which we have simultaneous measurements from all diagnostics. We note that the plasma quantities are not expected to vary significantly along the axial $z$-direction. The experimental results are recorded at 170~ns after current start. The sonic Mach number is calculated with $\gamma = 5/3$.}
\begin{tabular}{ccc}
\hline
Quantity & Value \\
\hline
$n_e$  [$\SI{}{\per \centi \meter \cubed}$]& $\SI{4e17}{} \pm \SI{1e17}{}$  \\
$T_e$ [$\SI{}{\electronvolt}$] & 11 \text{-} 14.5 \\
$T_i$ [$\SI{}{\electronvolt}$] &  6 \text{-} 9 \\
$V$ [$\SI{}{\kilo\meter\per\second}$] & 105 \text{-} 117  \\
$B$ [$\SI{}{\tesla}$] & \SI{2.2}{} \\
$n_i = n_e/\bar{Z}$  [$\SI{}{\per \centi \meter \cubed}$]& $\SI{1e17}{} \pm \SI{0.25e17}{}$  \\
$p = n_eT_e + n_i T_i$ [$\SI{}{\mega\pascal}$]& 0.6 \text{-} 1.1 \\
$\beta = p / (B^2/2\mu_0)$ [-]& 0.3 \text{-} 0.5 \\
$M_S = V / C_S$ [-] & 5.3 \text{-} 7.7 \\
$R_m = VL/\bar{\eta}$ [-] & 23 \\
\hline
\hline
\end{tabular}
\label{tab:table}
\end{table}

\section{Discussion}

\label{sec:discussion}

{\it Shock Geometry –} To determine the position of the shock front from the $\langle n_eL \rangle$ measurements shown in \autoref{fig:neL_lineouts}, we use an automated algorithm, where we fit a sigmoid function to the $\langle n_eL \rangle$ transition at the shock, and define the location with the largest gradient in $\langle n_eL \rangle$ as the shock front position. The vertical lines in \autoref{fig:neL_lineouts}\blue{a} show the detected shock front using this method, while the shaded regions show the associated uncertainty [here, equal to $0.5\times$FWHM (full width at half maximum) of the transition]. The dashed lines in \autoref{fig:neL_lineouts}\blue{b} and \autoref{fig:neL_lineouts}\blue{c} show the detected shock front at the lower obstacle overlaid on the interferogram and shadowgraph respectively. Although the shadowgraph was not used to determine the shock front location, the detected shock front matches the start of the dark band seen in the shadowgraph well, consistent with the sharp electron density gradients that deflect the probing beam away from the shock front. The detected shock front at the upper obstacle, obtained using the same methodology, is shown in \autoref{fig:neL_lineouts_top}. The shock fronts make angles of about $\SI{14}{\degree}\pm\SI{0.5}{\degree}$ and $\SI{32}{\degree}\pm\SI{0.5}{\degree}$ with respect to the horizontal.

{\it Density Jump –} The $\langle n_e L\rangle$ measurements can also help us constrain the compression ratio across the shock front. We estimate the jump in electron density across the shock using the expression  $\bar{n}_{e,2}/\bar{n}_{e,1} \approx 1 + \left( \langle n_eL \rangle_2 - \langle n_eL \rangle_1 \right)/(2l\bar{n}_{e,0})$, which is derived in \autoref{appendixB}. Here, $\langle n_eL \rangle_{2,1}$ are the post- and pre-shock values of the line-integrated electron density measured using interferometry, $2l = \SI{3.0}{\milli\meter}$ is the width of the obstacle in the $y$-direction, and $\bar{n}_{e,0}$ is the pre-shock electron density. Using this expression, for the bottom shock at $x = \SI{8}{\milli\meter}$, we find that the jump in electron density is about $\bar{n}_{e}/\bar{n}_{e,0} \approx 18$.  

We estimate the post-shock ionization using the SpK nLTE model \citep{crilly2023spk}.  The upper bound on the post shock-ionization, assuming a temperature jump of $T/T_0 \approx 11$ from the (hydrodynamic) normal shock relations (for $M_0 \approx 5.6, \, \gamma = 5/3$), is $\bar{Z} \approx 11$, roughly $2.75\times$ larger than the pre-shock ionization $\bar{Z}_0 \approx 3.9$. Therefore, the compression of the plasma density must be at least $r_1 \equiv n_{i}/n_{i,0} \gtrsim 7$. \black{As described earlier, we are unable to fully resolve the change in $\langle n_eL \rangle$ across the top shock at $\SI{8}{\milli\meter}$, because the fringes become too close to trace properly (see \autoref{fig:neL_lineouts_top}). Further away at $x = \SI{15}{\milli\meter}$, however, we expect a similarly large jump in $\bar{n}_e$ of about $\bar{n}_{e}/\bar{n}_{e,0} > 30$ to account for the change in $\langle n_eL \rangle$ measured at this location.}

\begin{figure*}[t!]
\includegraphics[page=9,width=1.0\textwidth]{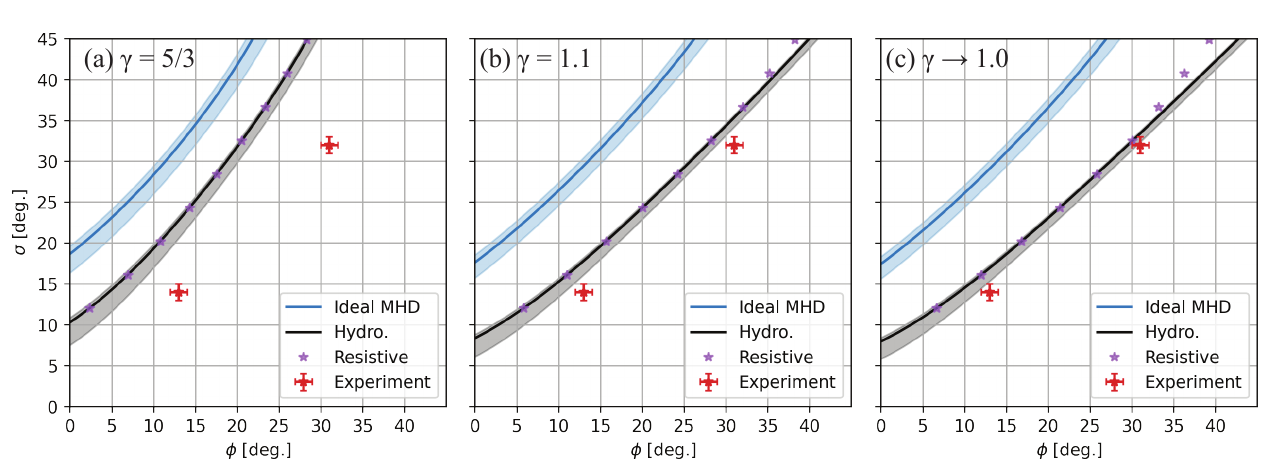}
\centering
\caption{ Variation of the shock angle $\sigma$ with the deflection angle $\phi$ for the experimentally measured pre-shock plasma conditions (see \autoref{tab:table}), using the ideal MHD (\autoref{eq:perp_eqn}) [blue] and hydrodynamic (\autoref{eq:phi_vs_sigma_hydro}) [black] solutions. The shaded regions represent the uncertainty in the experimental measurements, while the red markers represent the angles observed in the experiment. We show results for three different values of the polytropic index $\gamma = 5/3, \, 1.1, \text{ and } 1$. The purple markers represent the solution caluclated using the resistive model with $f = \text{min}\left[(M_A R_M \sin^2\sigma)^{-1},f_{max}\right]$.}
\label{fig:comparison}
\end{figure*}

{\it Resistive and Cooling Effects –}  For the experimentally measured quantities listed in \autoref{tab:table}, we expect both resistive diffusion and radiative cooling to be important in our system. We estimate a magnetic Reynolds number of $R_M \approx 23$, calculated using Spitzer resistivity and a system-size length scale $L = \SI{1}{\centi \meter}$, which is similar to the value obtained in previous pulsed-power-driven shock experiments \citep{Burdiak2017,datta2022structure,russell2022perpendicular}. The resistive diffusion length is $l_\eta \sim \bar{\eta}/V \approx \SI{0.5}{\milli\meter}$, which is on the order of the shock stand-off observed in the experiment (in other words, $R_M \approx 2$ when calculated using the width of the shocked plasma $\sim \SI{1}{\milli\meter}$), indicating that resistive effects can smooth large gradients in the magnetic field across the shock front, as described earlier in \blue{Sec.}~\ref{sec:theory}. 

To quantify the importance of radiative cooling, we require an estimate of the post-shock temperature and pressure. For $M_0 = 5.6, \gamma = 5/3$, and using the purely hydrodynamic oblique shock relations for simplicity, the temperature change is expected to be about $T/T_0 \approx 2-6$ for the given deflection angles. An estimate of the radiative cooling time can be obtained from $\tau_\text{cool} \equiv [p/(\gamma -1)]/q_\text{rad}$, where $p$ is the post-shock pressure, and $q_\text{rad}$ is the volumetric radiative heat loss, calculated using the post-shock density and temperature. Using the temperature jump calculated from the oblique shock relations, and $q_\text{rad}$ computed using the nLTE SpK model (which includes line, recombination, and bremsstrahlung losses) \citep{crilly2023spk}, we expect the cooling time to be $\tau_\text{cool} \approx 0.2\text{-}\SI{7}{\nano\second}$, which is short when compared to the hydrodynamic time $L/V \approx \SI{100}{\nano\second}$, indicating that cooling effects may be significant. Note that we include optical depth effects in calculating the radiative cooling rate here, by solving radiation transport along a planar path of length equal to the half-width $l = \SI{1.5}{\milli\meter}$ of the obstacle, using nLTE emissivity and opacity data generated by SpK \citep{datta2024radiatively}. In the experiment, although not shown here, XUV and optical emission from the system was also recorded, and the shocked plasma indeed generated significant radiative emission compared to the pre-shocked plasma. 

For characteristic values of $n_i = 1-\SI{10e17}{\per \centi\meter\cubed}$ and $T_e \approx 10-\SI{20}{\electronvolt}$, the SpK results indicate that the aluminum emission spectrum is dominated by XUV/soft X-ray emission, peaking around photon energies of $\sim \SI{100}{\electronvolt}$ ($\lambda \approx \SI{12}{\nano\meter}$). This is primarily due to strong line radiation from the partially ionized aluminum plasma, which causes the total emission to significantly exceed bremsstrahlung and recombination contributions in this temperature range \citep{datta2024simulations}. The spectral opacity is also maximum around photon energies of $\hbar\omega \approx \SI{100}{\electronvolt}$. For the post-shock plasma with characteristic density $n_i \sim \SI{1e18}{\per\centi\meter\cubed}$, the characteristic absorption length (estimated as the reciprocal of the spectral opacity $\alpha_\omega$) is $\alpha_\omega ^{-1}\approx 1-\SI{3}{\milli\meter}$ for photon energies around $\SI{100}{\electronvolt}$, while for lower photon energies in the UV and optical ranges, the absorption length is $>\SI{30}{\milli\meter}$. Since the length scale of the shocked plasma is $l =\SI{1.5}{\milli\meter}$, we expect a significant portion of the emitted radiation to escape, rather than be re-absorbed by the plasma. The pre-shock plasma, with characteristic density $n_i \sim \SI{1e17}{\per\centi\meter\cubed}$, temperature $T_e \approx \SI{10}{\electronvolt}$ and length scale $\sim \SI{10}{\milli\meter}$, is also expected to be mostly optically thin – the minimum absorption length is $1/\alpha_\omega \approx \SI{10}{\milli\meter}$ (for $\hbar\omega \approx \SI{100}{\electronvolt}$), while that for lower photon energies is large $>\SI{1}{\meter}$. Therefore, although some re-absorption of radiation may occur, especially in the photon energy range around \SI{100}{\electronvolt}, we expect a significant portion of the emitted radiation to escape from the system, contributing to cooling of the plasma. This justifies the thin-thin radiative shock approximation used in \autoref{sec:theory}.

{\it Comparison with analytical models –} Finally, we compare our experimental results to the theoretical models described in \blue{Sec.}~\ref{sec:theory}. As shown earlier, while an out-of-plane magnetic field increases the shock angle for a given deflection angle, resistive diffusion and radiative cooling can reduce the shock angle. In \autoref{fig:comparison}, we plot the variation of the expected shock angle $\sigma$ with the deflection angle $\phi$ for the experimentally measured pre-shock plasma conditions (see \autoref{tab:table}), using the ideal MHD (\autoref{eq:perp_eqn}),  and hydrodynamic (\autoref{eq:phi_vs_sigma_hydro}) solutions. The shaded regions represent the range of the experimental measurements, while the red markers represent the angles observed in the experiment. We show results for three different values of the polytropic index $\gamma = 5/3, 1.1, \text{ and } 1$. The value $\gamma \approx 1.1$ represents the predicted magnitude of the polytropic index that accounts for deviations from ideal gas behavior in a high energy density plasma \citep{Drake2006}. Similarly, the value $\gamma \rightarrow 1$ result serves as a proxy for the effects of radiative cooling, which as mentioned before describes an idealized system with equal pre- and post-shock temperatures \citep{lacey1988structure}.


The ideal MHD result overpredicts the shock angle for all values of the polytropic index $1 < \gamma < 5/3$. This indicates that a departure from ideal MHD behavior is likely, causing the result to be closer to the hydrodynamic solution. Indeed, as noted earlier, we expect magnetic diffusion to be important, given that the resistive diffusion length is comparable to the length scale of the shocked plasma. A very rough estimate of the dimensionless magnetic transport $f$ across the shock front can be obtained as $f \sim (\bar{\eta}B/l)/(u_{n,0}^2\sqrt{\mu_0\rho_0}) \approx (M_{A,n} R_M \sin \sigma)^{-1}$. Here, $R_M = u_0l/\bar{\eta}$ and $l$ is the relevant length scale of shocked plasma, which is distinct from the system size length scale $L = \SI{1}{\centi\meter}$ used to calculate $R_M$ in \autoref{tab:table}. We plot the solution for the shock angle using $R_M \approx 2$ and setting $f = \text{min}\left[(M_A R_M \sin^2\sigma)^{-1},f_{max}\right]$, and find that the resistive solution (\blue{Equations \ref{eq:final_eqn} $\&$ \ref{eq:phi_vs_sigma})} is expected to be close to the hydrodynamic solution, as shown in \autoref{fig:comparison}. 

As we decrease $\gamma$, the theoretical prediction of the hydrodynamic model becomes closer to the experimental values; however, for $\gamma = 1.1$ the theoretical shock angle still exceeds the experimental measurement (see \autoref{fig:comparison}\blue{b}). The theoretical prediction is closest to the experimental result for $\gamma \rightarrow 1$, which would indicate that there is significant cooling of the post-shock plasma, and that a reduction in the polytropic index $\gamma$ due to HED effects alone cannot account for the shallow angles seen in the experiment. The predicted shock angles for $\gamma \rightarrow 1$ are $\sigma_1 \approx \SI{16.3}{\degree}\pm \SI{1}{\degree}$, and $\sigma_2 \approx \SI{32.9}{\degree}\pm \SI{0.7}{\degree}$, slightly higher than the experimental values, while the predicted density compression across the shock ($r_1 \approx 5.3 \pm 1.7$, $r_2 \approx 21 \pm 7$), are lower than the experimental estimate. 

A possible mechanism that may further reduce the shock angle is 3-D pressure relief \citep{Anderson1990}. Although an analytic result for the shock angle under 3-D pressure relieving effects is challenging to obtain, we can estimate a lower bound from the solution of the Taylor-Maccoll equation, which describes shock formation at a conical obstacle \citep{Anderson1990}. For $\gamma = 1.1$ and a cone angle $\SI{13}{\degree}$, we expect the shock angle to decrease from the 2-D oblique shock result $\SI{18}{\degree}$ to $\SI{16}{\degree}$ (for the hydrodynamic solution). Similarly, for a cone angle $\SI{31}{\degree}$ and $\gamma = 1.1$, the conical shock angle is about $\SI{33}{\degree}$, as opposed to the the 2-D oblique shock result of about $\SI{35}{\degree}$. Although some 3-D reliving may occur in the experiment, since the conical shock angle solutions are still higher than the experimentally observed values, this indicates that 3-D effects alone cannot account for the shallow shock angles in the experiment. Furthermore, 3-D effects would further reduce the post-shock compression, and thus not be consistent with the large density jump seen in the experiment. 

\begin{figure}[t!]
\includegraphics[page=10,width=0.48\textwidth]{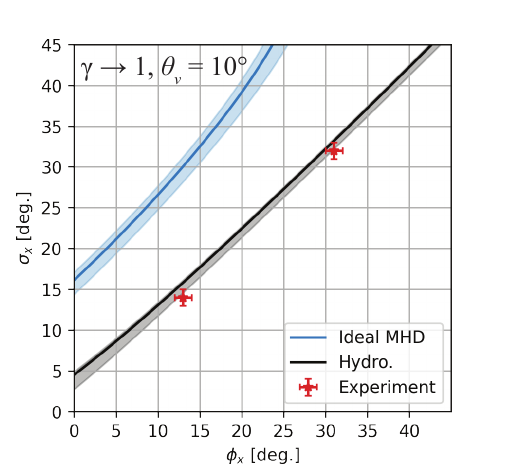}
\centering
\caption{The shock angle as a function of the deflection angle for the experimentally measured pre-shock plasma conditions (see \autoref{tab:table}), using the ideal MHD (\autoref{eq:perp_eqn}) [blue], and hydrodynamic (\autoref{eq:phi_vs_sigma_hydro}) [black] solutions. The results are shown for $\gamma = 1$ and $\theta_v = \SI{10}{\degree}$. We show the shock and deflection angles measured relative to the $x$-axis.
}
\label{fig:velocity_angle}
\end{figure}

In the analysis shown in \autoref{fig:comparison}, we have assumed an upstream velocity directed along the $x$-direction. As described earlier, expansion of the plasma in the $z$-direction can result in a small but non-zero $z$-component of the velocity \citep{datta2022structure}. A rough estimate of this expansion velocity is provided by $V_z \sim \sqrt{p/\rho}$, which is about $\SI{16}{\kilo\meter\per\second}$, providing a velocity angle of $\theta_v = \tan^{-1}(V_z/V_x)\approx \tan^{-1}(16/100) \approx \SI{9}{\degree}$. This suggests that a significant $z$-component of velocity can exist upstream of the shock. We repeat the previous analysis, but with a velocity angle $\theta_v = \SI{10}{\degree}$ directed clockwise from the horizontal for the bottom obstacle, and counter-clockwise for the top. The expected shock and deflection angles with respect to the horizontal are shown in \autoref{fig:velocity_angle}. The actual shock and deflection angles should be calculated with respect to the upstream velocity vector, that is, $\sigma = \sigma_x + \theta_v$ and $\phi = \phi_x + \theta_v$. 

As shown in \autoref{fig:velocity_angle}, accounting for the upstream velocity direction provides a better agreement with the experimentally observed shock angles. Furthermore, the density compression is predicted to be $r_1 \approx 12 \pm 4$ and $r_1 \approx 33 \pm 12$ for the bottom and top shocks respectively, closer to the large values measured in the experiment. Finally, we note that even for large values of $\theta_v$, an agreement with the experimentally observed shock angles is not obtained for the ideal MHD model. The same is true for the hydrodynamic model if $\gamma \neq 1$. This suggests that the shock is best represented by a system dominated by both radiative cooling and resistive diffusion. This result is consistent the predicted values of the resistive diffusion length $l_\eta \approx \SI{0.5}{\milli\meter}$ and the radiative cooling times $\tau_\text{cool} \approx 0.2\text{-}\SI{7}{\nano\second}$ listed earlier. \black{In the present experiments, diagnostic limitations preclude direct measurements of the change in the magnetic field, temperature, and the velocity vector across the shock. Although our analysis indicates the importance of resistive diffusion, cooling, and velocity direction, future experiments that directly measure post-shock quantities can help confirm these effects. Furthermore, we have assumed for simplicity that the quantities measured by our diagnostics do not change significantly along the $z$-direction. While this is typically true for wire arrays, the transport of magnetic field ahead of the shock may affect the momentum and energy balance, altering pre-shock conditions.}

\textit{Electrothermal Instability -} We now return to the small scale perturbations seen in the shadowgraphs (see \autoref{fig:neL_lineouts} and \autoref{fig:neL_lineouts_top}). We hypothesize that these are caused by the electrothermal instability (ETI). The surface modulations exhibit a wavevector $k$ parallel to the current density, which is consistent with the striation form of ETI, excited on conducting surfaces with $d\eta/dT>0$ \citep{peterson2012electrothermal}. Assuming an angular frequency $2\pi/(4\times \SI{120}{\nano\second})$ and aluminum resistivity $\eta \approx \SI{2.8e-8}{\ohm \meter}$ gives a skin depth $\delta \approx \SI{0.06}{\milli\meter}$, which combined with the maximum current of $I_{max} \approx \SI{1}{\mega\ampere}$ provides a skin current density of $ J \approx I_{max}/(\delta  w) \approx \SI{5e12}{Am^{-2}}$. A rough estimate of the upper bound on the ETI growth rate, assuming values of mass density $\rho = \SI{2700}{\kilo\gram\cubic \meter}$, specific heat $c_v = \SI{927}{\joule \per \kilo \gram \per \kelvin}$, thermal conductivity $\kappa \approx \SI{200}{\watt\per\meter\kelvin}$, and $d\eta/dT \sim 1 \times 10^{-10} \Omega \mathrm{~m} \mathrm{~K}^{-1}$ is then $\gamma=[(d\eta/dT) J^2-k^2 \kappa] / (\rho c_v) \approx \SI{1.3}{\per \nano \second}$ \citep{peterson2012electrothermal}. Thus, ETI growth on the obstacle surface is plausible on the experimental time scale.

\section{Conclusions and Future Work}

We present experimental results describing magnetized oblique shock formation in a plasma where resistive diffusion and radiative cooling are important. In \blue{Sec.}~\ref{sec:theory}, we outline a theoretical model that describes oblique shock formation in a magnetohydrodynamic (MHD) fluid with an out-of-plane magnetic field. In the ideal MHD limit, the magnetized shock forms at a larger angle than in the hydrodynamic limit. However, we show that strong resistive diffusion and radiative cooling can reduce the shock angle (see \autoref{fig:obl_model}).

In the experiment, magnetized plasma flows ($M_1 \approx 5.6, \beta \approx 0.5$) generated by an exploding wire array are deflected by two planar obstacles placed at $\phi_{1,2} = \SI{13}{\degree}, \SI{31}{\degree}$ with respect to the horizontal. The interaction of the supersonic, super-Alfvénic inflows with the planar obstacle generates oblique shocks, which are visualized using Mach-Zehnder interferometry (\autoref{fig:results}\blue{d}) and laser shadowgraphy (\autoref{fig:results}\blue{f}), while the pre-shock plasma conditions (velocity, ionization, ion and electron temperatures, and magnetic field) are measured using optical Thomson scattering (OTS) (\autoref{fig:OTS_result}) and B-dot probes (\autoref{fig:results}\blue{b}). The shocks  form at shallow angles of about $\SI{14}{\degree}\pm\SI{0.5}{\degree}$ and $\SI{32}{\degree}\pm\SI{0.5}{\degree}$ with respect to the horizontal. The shocks also exhibit large electron density jumps across the shock front.

In the experiment, the resistive diffusion length $l_\eta \approx \SI{0.5}{\milli\meter}$ is comparable to the length scale of the shocked plasma, and the radiative cooling time $\tau_\text{cool} \approx 0.2\text{-}\SI{7}{\nano\second}$ is expected to be small when compared to the hydrodynamic time $V/L \approx \SI{100}{\nano\second}$, indicating that resistive diffusion and radiative cooling are important. A comparison of the experimentally observed shock angles and density compression with the theoretical predictions support this result. As shown in \autoref{fig:comparison}, ideal MHD theory cannot account for the shallow angles observed in the experiment. The best agreement is found for the hydrodynamic oblique shock model with a $\gamma_\text{eff} \rightarrow 1$, when a small non-zero velocity angle $\theta_v \approx \SI{10}{\degree}$ is assumed (see \autoref{fig:velocity_angle}), providing a closer match to both shock angles, and the density compression observed experimentally. 

In the present experiments, the proximity of the shock to the obstacle precluded direct measurements of the change in the magnetic field, temperature, and the velocity vector across the shock. In future experiments, a tuning of the wire array properties, and hence of the pre-shock plasma, could be pursued to explore oblique shock formation at wider angles. Another important measurement would be the spatial variation in the magnetic field, which could be recorded through Faraday rotation. In the present set of experiments, this diagnostic was unavailable, but could be useful in future experiments to visualize the transport of magnetic field ahead of the shock. Finally, the present platform also demonstrates evidence for ETI-like perturbations on the obstacle surface. Although ETI-characterization is not the focus of the present study, this platform could be used to study ETI growth with plasma flow directed along the conducting surface, which is of potential interest in MagLIF studies.

\section{Acknowledgments}
The authors thank Todd Blanchard, Dan Hawkes, and Harry Wilhelm for their work in support of the experiments. Funding for this research was provided by the National Nuclear Security Administration through the ZNetUS program, and was also supported in part by the NNSA Stewardship Science Academic Programs under DOE Cooperative Agreement DE-NA0004148.
\section{Declaration of Conflicts of Interest}

The authors have no conflicts of interest to disclose.

\section{Data Availability}

The data that support the findings of this study are available from the corresponding author upon reasonable request.

\begin{appendices}

\section{Appendix A - Numerical Results}
\label{appendixA}

\setcounter{equation}{0}
\numberwithin{equation}{section} 

To demonstrate the analytical result shown in \autoref{fig:obl_model}\blue{b}, we perform two-dimensional ($xz$-plane) nonlinear MHD simulations. The compressible MHD equations are solved using the MacCormack scheme, which is an explicit second-order finite-difference shock capturing method \citep{goedbloed_keppens_poedts_2010}. We use a uniform $256 \times 512$ rectilinear grid with a grid resolution of $\Delta x = 1.0/256$, unless otherwise specified. A convergence study showed that this resolution was sufficient for convergence. Adaptive time stepping based on the Courant–Friedrichs–Lewy condition is used, and a small amount of artificial viscosity is used to ensure numerical stability. As an example, we initialize the simulation with uniform density $\rho_0 = 1.0$, pressure $p_0 = 1.0$, and velocity $u_0 = 3.62$. The quantities are represented in non-dimensional units. A constant $\gamma = 5/3$ is used in the simulations. The upstream Mach number is therefore $M_0 = 2.81$. The out-of-plane magnetic field $B_0 = 1.0$. Here, we set $\mu_0 = 1$, which gives a plasma $\beta = 2$. The flow is deflected by a perfectly conducting wedge-shaped obstacle, providing a deflection angle of $\phi = \SI{10}{\degree}$. A no flux ${\bf v \cdot n} = 0$ boundary condition is prescribed on the wedge surface, such that the velocity can only be tangential along the wedge. No gradient boundary conditions are prescribed at the boundaries of the simulation domain.  Each simulation is run until a steady-state solution is obtained, which happens within a few hydrodynamic times $\sim \tau_H \equiv 1.0/u_0 \approx 0.27$.

\begin{figure}[t!]
\includegraphics[page=2,width=0.48\textwidth]{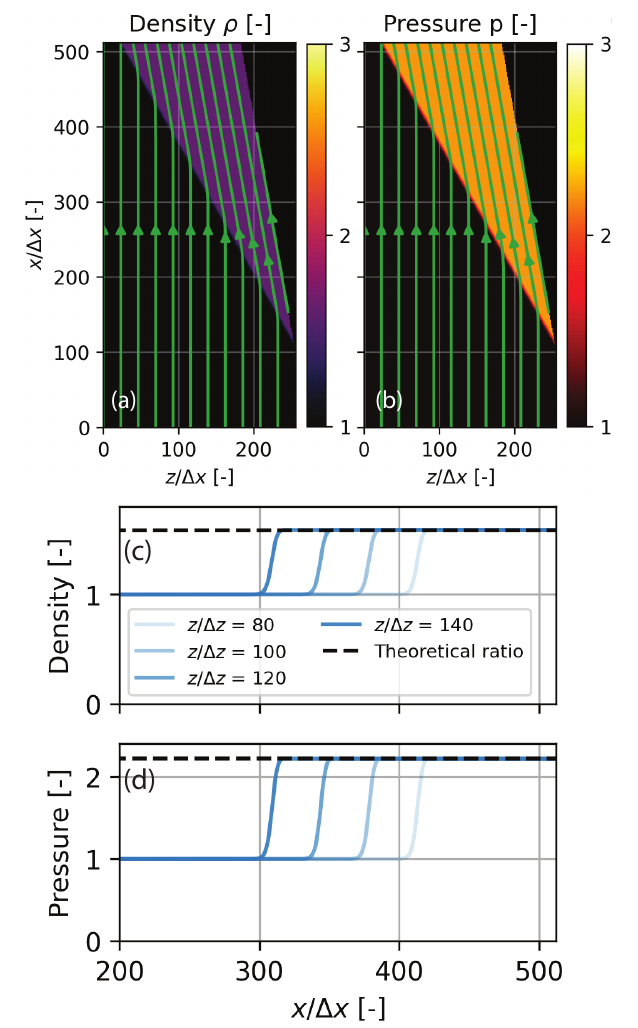}
\centering
\caption{ (a-b) Density and pressure for the hydrodynamic case. The wedge angle is $\phi = \SI{10}{\degree}$, $\gamma = 5/3$, and the upstream Mach number is $M_0 = 2.81$. (c-d) Lineouts of the density and pressure along $x$ for different $z$-positions. Analytical predictions of the compression ratios are shown using the horizontal dashed black lines.
}
\label{fig:sim_hydro}
\end{figure}

\begin{figure}[t!]
\includegraphics[page=3,width=0.48\textwidth]{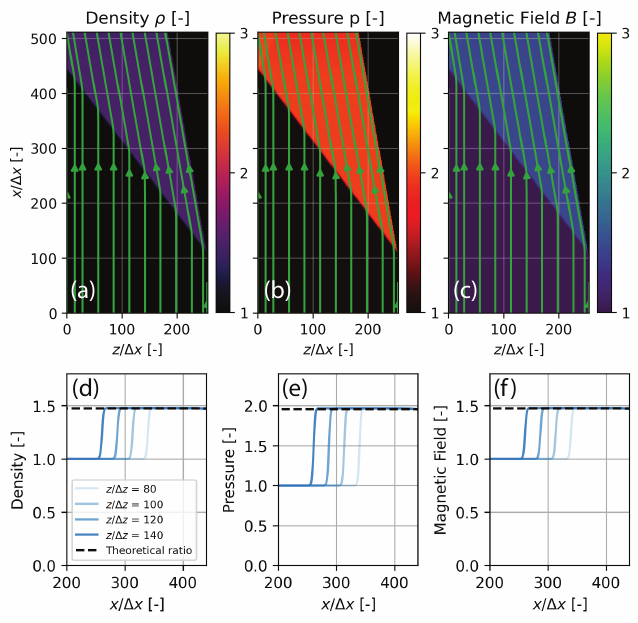}
\centering
\caption{ (a-c) Density, pressure, and magnetic field for the $\beta = 2, \gamma = 5/3, M_0 = 2.81$ MHD shock. The wedge angle is $\phi = \SI{10}{\degree}$. (d-f) Lineouts of the density, pressure, and magnetic field along $x$ for different $z$-positions. Analytical predictions of the compression ratios are shown using the horizontal dashed black lines.
}
\label{fig:sim_mhd}
\end{figure}

\begin{figure}[b!]
\includegraphics[page=4,width=0.48\textwidth]{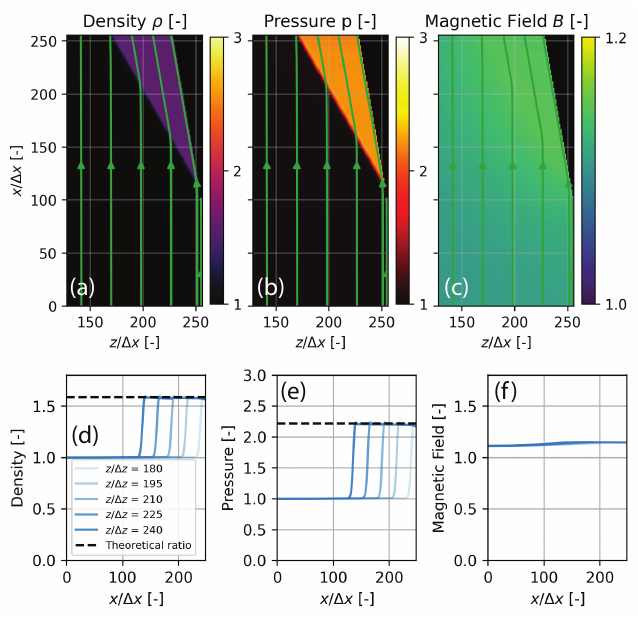}
\centering
\caption{ (a-c) Density, pressure, and magnetic field for the $\beta = 2, R_M \approx 1, M_0 = 2.81$ resistive oblique shock. The wedge angle is $\phi = \SI{10}{\degree}$. (d-f) Lineouts of the density, pressure, and magnetic field along $x$ for different $z$-positions. Analytical predictions of the compression ratios are shown using the horizontal dashed black lines.
}
\label{fig:sim_resistive}
\end{figure}

{\textit{Hydrodynamic Result –}} First, we show the the purely hydrodynamic case, where the magnetic field is $B = 0$. \autoref{fig:sim_hydro}\blue{(a-b)} show the distribution  of density and pressure for this case, together with the velocity streamlines. The shock forms at $\sigma =\SI{30}{\degree}$, consistent with the theoretical prediction shown in \autoref{fig:obl_model}\blue{b}. Lineouts of the density and pressure along $x$ for different $z$-positions are shown \autoref{fig:sim_hydro}\blue{(c-d)}. The density and pressure compression ratios are $r = 1.59$ and $R = 2.22$, consistent with the values expected from analytical hydrodynamic theory; the analytical predictions of the compression ratios are shown using the horizontal dashed black lines in the \autoref{fig:sim_hydro}\blue{(c-d)}.

{\textit{Ideal MHD Result –}} The ideal MHD result for $B_0 = 1.0$ ($\beta = 2$) is shown in \autoref{fig:sim_mhd}.  The shock forms at a higher angle $\sigma =\SI{37}{\degree}$ than in the hydrodynamic case, and the shock angle agrees with the theoretical prediction shown in \autoref{fig:obl_model}\blue{b}. Lineouts of the density, pressure, and magnetic field along $x$ for different $z$-positions are shown \autoref{fig:sim_mhd}\blue{(d-f)}. The density and magnetic field exhibit equal compression ($r = 1.47$), consistent with the theoretical prediction shown using the dashed black line. The pressure compression is $R = 1.96$, again consistent with the value expected from analytical theory.

\begin{figure*}[t!]
\includegraphics[page=5,width=0.98\textwidth]{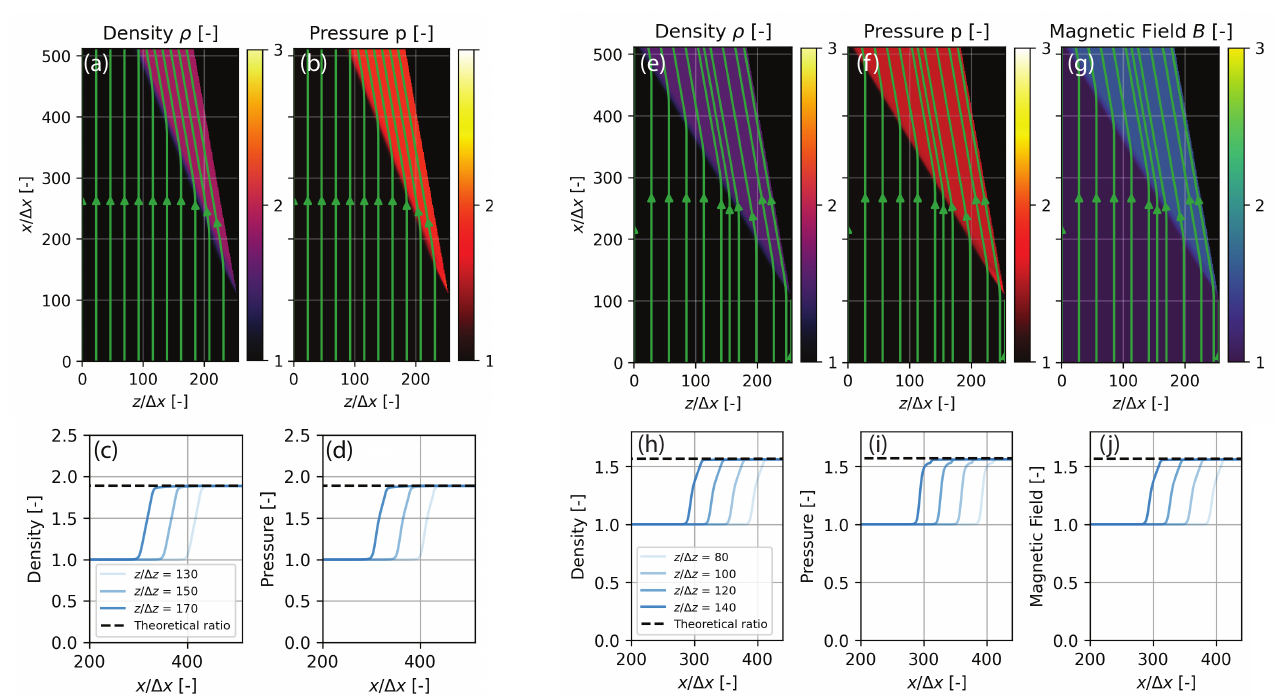}
\centering
\caption{ Radiatively cooled oblique shock. The wedge angle is $\phi = \SI{10}{\degree}$. The upstream Mach number is $M_0 = 2.81$, and $\gamma = 5/3$. (a) and (b) show the steady state density and pressure in the hydrodynamic case. (c) and (d) show lineouts of density and pressure along x for different $z$-positions. (e), (f) and (g) show the steady state density, pressure, and magnetic field in the $\beta = 2$ MHD case. (h), (i), and (j) show lineouts of density, pressure, and magnetic field along x for different $z$-positions.
}
\label{fig:sim_cooling}
\end{figure*}

{\textit{Resistive MHD Result –}} We repeat the previous simulation ($M_0 = 2.81, \beta = 2, \phi =\SI{10}{\degree}$) with a large value of magnetic diffusivity, providing a magnetic Reynolds number $R_M = u_0L/\bar{\eta} \approx 1$. \autoref{fig:sim_resistive}\blue{(a-c)} show the density, pressure, and magnetic field respectively. The resistivity is held constant throughout the simulation. The density and pressure jumps are sharp, while the magnetic field is comparatively smooth, and a magnetic precursor extends beyond the shock. The shock now forms at $\sigma = \SI{30}{\degree}$, consistent with the purely hydrodynamic result.  Lineouts of the density, pressure, and magnetic field along $x$ for different $z$-positions are shown \autoref{fig:sim_hydro}\blue{(c-d)}. The density and pressure compression ratios are $r = 1.59$ and $R = 2.22$, identical to the hydrodynamic result (\autoref{fig:sim_hydro}). 

{\textit{Radiative Cooling –}} We repeat the hydrodynamic and ideal MHD simulations with a volumetric cooling term in the energy equation $q_\text{rad}$. The volumetric cooling rate in set to 0 for $T \equiv p / \rho \leq 1$. This is done to ensure that the pre-shock conditions do not change during the simulation. The system is therefore driven to thermal equilibrium by radiative cooling, resulting in an isothermal system with equal pre- and post-shock temperatures. We use a bremsstrahlung-like volumetric cooling term $q_\text{rad} = A\rho^2T^{1/2}$ in the simulations for $T > 1$, and the non-dimensional cooling rate in the post-shock region is about $\tau_\text{cool}^{-1}/\tau_H^{-1} \approx 2.5$. For simplicity,  the resistivity is held constant throughout the simulations. The initial conditions in these simulations are the non-radiative ideal MHD solutions described previously (\autoref{fig:sim_hydro} for the hydrodynamic case, and \autoref{fig:sim_mhd} for the MHD case). 

With radiative cooling, the shock angle in the simulations decreases until a steady solution is reached. The temperature rises at the shock front but then falls rapidly behind it due to radiative cooling. The solution for the hydrodynamic case is shown in \autoref{fig:sim_cooling}\blue{(a-d)}, while that for the ideal MHD case is shown in \autoref{fig:sim_cooling}\blue{(e-j)}. In both cases, the shock angle is lower than in the non-radiative cases. The hydrodynamic simulation demonstrates a shock angle of about $\sigma \approx \SI{22}{\degree}$, with equal compression ratios for the density and pressure $r = R = 1.9$. Similarly, the MHD simulation exhibits a shock angle of about $\sigma \approx \SI{31.5}{\degree}$, and equal compression ratios for the density, magnetic field, and pressure $r = R = 1.57$.  The dashed black lines in \autoref{fig:sim_cooling}\blue{(c-d)} and \autoref{fig:sim_cooling}\blue{(h-j)} represent the analytical compression ratios obtained for $\gamma_\text{eff} \rightarrow 1$ and $M_\text{eff} = u_0 / \sqrt{p_0/\rho_0} \approx 3.62$, showing a good match to the simulated solution. The simulated shock angles also agree with the analytical result of \autoref{eq:phi_vs_sigma}.

\section{Appendix B - Electron Density Jump}
\label{appendixB}

\setcounter{equation}{0}
\numberwithin{equation}{section} 

The line-integrated electron density at a given position $x = x_0$ in the $xz$-plane, assuming probing along the $y$-direction is:

\begin{equation}
\langle n_e L \rangle_{x_0} = \int_{-Y}^{Y} n_e(y,x_0) dy
\end{equation}

Here, $|y|\leq Y$ is the extent of the electron density distribution along $y$. Assuming that the shocked plasma is localized to $|y|\leq l$, we can rewrite the above expression as:

\begin{equation}
\langle n_e L \rangle_{x_0} = \int_{-Y}^{-l} n_e(y,x_0) dy + \int_{-l}^{l} n_e(y,x_0) dy + \int_{l}^{Y} n_e(y,x_0) dy
\label{eq:postshock_den}
\end{equation}

Similarly, for the density distribution $n_e(x,y)^*$ in the $xz$-plane without a shock, we can write a similar expression:

\begin{equation}
\langle n_e L \rangle_{x_0}^* = \int_{-Y}^{-l} n_e^*(y,x_0) dy + \int_{-l}^{l} n_e^*(y,x_0) dy + \int_{l}^{Y} n_e^*(y,x_0) dy
\end{equation}

Re-writing $\int_{-l}^{l} n_e^*(y,x_0) dy \approx \bar{n}_{e,1} (2l)$, where $\bar{n}_{e,1}$ is the pre-shock electron density, we get:

\begin{equation}
\langle n_e L \rangle_{x_0}^* - \bar{n}_{e,1} (2l) = \int_{-Y}^{-l} n_e^*(y,x_0) dy + \int_{l}^{Y} n_e^*(y,x_0) dy
\label{eq:preshock_den}
\end{equation}

Going back to the plane containing the shock, we assume that for the region outside the shocked plasma, the densities are equal, that is, $n_e(y,x_0) = n_e^*(y,x_0) \text{ for } |y|> l$. We can then re-write \autoref{eq:postshock_den} as:

\begin{equation}
\langle n_e L \rangle_{x_0} =  \bar{n}_{e,2}(2l) + \langle n_e L \rangle_{x_0}^* - \bar{n}_{e,1} (2l)
\end{equation}

Here, we have substituted in \autoref{eq:preshock_den} and further assumed $\int_{-l}^{l} n_e(y,x_0) dy \approx \bar{n}_{e,2}(2l)$, where $\bar{n}_{e,2}$ is the post-shock electron density. The above expression can finally be re-written in terms of the electron density jump:

\begin{equation}
\frac{\bar{n}_{e,2}}{\bar{n}_{e,1}} = 1 + \frac{\langle n_e L \rangle_{x_0} - \langle n_e L \rangle_{x_0}^*}{2l\bar{n}_{e,1}}
\end{equation}

Here, $\langle n_e L \rangle_{x_0}$ and $\langle n_e L \rangle_{x_0}^*$ are the post-shock and pre-shock line-integrated electron densities measured using interferometry, $2l = \SI{3}{\milli\meter}$ is set by the width of the obstacle, and we estimate $\bar{n}_{e,1}$ using an Abel inversion.

\end{appendices}

\bibliography{main}

\end{document}